# REINSURING AI:
## ENERGY, AGRICULTURE, FINANCE & MEDICINE AS PRECEDENTS FOR SCALABLE GOVERNANCE OF FRONTIER ARTIFICIAL INTELLIGENCE

*Nicholas Stetler[1]*

## ABSTRACT

The governance of frontier artificial intelligence (AI) systems—particularly those capable of catastrophic misuse or systemic failure—requires institutional structures that are robust, adaptive, and innovation-preserving. This paper proposes a novel framework for governing such high-stakes models through a three-tiered insurance architecture: (1) mandatory private liability insurance for frontier model developers; (2) an industry-administered risk pool to absorb recurring, non-catastrophic losses; and (3) federally backed reinsurance for tail-risk events.

Drawing from historical precedents in nuclear energy (Price-Anderson), terrorism risk (TRIA), agricultural crop insurance, flood reinsurance, and medical malpractice, the proposal shows how the federal government can stabilize private AI insurance markets without resorting to brittle regulation or predictive licensing regimes. The structure aligns incentives between AI developers and downstream stakeholders, transforms safety practices into insurable standards, and enables modular oversight through adaptive eligibility criteria.

By focusing on risk-transfer mechanisms rather than prescriptive rules, this framework seeks to render AI safety a structural feature of the innovation ecosystem itself—integrated into capital markets, not external to them. The paper concludes with a legal and administrative feasibility analysis, proposing avenues for statutory authorization and agency placement within existing federal structures.

---

[1] JD Candidate 2026, Southern Illinois University

**TABLE OF CONTENTS**

# INTRODUCTION

The beauty of mathematics is that its truths can be confirmed. A child, if brilliant enough, can outclass even their elders—as Terence Tao did, at the age of ten, when he won the 1986 International Math Olympiad.[2] But what happens when the frontier of reasoning is no longer represented by the human mind?

In late 2024, Tao and a team of mathematicians decided to give leading artificial intelligence labs the ultimate test.[3] They created *FrontierMath*, a benchmark of 300 unpublished problems designed to separate genuine abstract reasoning from AI's usual statistical tricks.[4] The goal was simple: see if today's best models could handle problems, demanding deep intuition, that often stump even professional mathematicians.[5]

It was *supposed* to be difficult. When asked about the difficulty of the test Tao stated: "These are extremely challenging… I think they will resist AIs for several years at least."[6] And yet, within weeks, Open AI's *o3* model solved more than 25% of them.[7] AI had not only passed the test—it had done so at a level that surprised even its creators.[8]

OpenAI's ambition is to develop artificial general intelligence (AGI)—systems capable of outperforming humans, not just at math, but at *everything*.[9] This aim has caused immense alarm among researchers and policymakers.[10] At the center of the concern lies the alignment problem: the difficulty of ensuring that powerful AI systems act in ways which reflect humankind's values, goals, and safety.[11] As OpenAI admits, "There is currently no known indefinitely scalable solution to the alignment problem. As AI progresses, we expect to encounter new challenges that we don't observe in current systems."[12]

---

[2] INT'L MATH. OLYMPIAD, *Terence Tao – IMO Official Results*, https://www.imo-official.org/participant_r.aspx?id=1581 (last visited Feb. 27, 2025).
[3] EPOCH AI, *FrontierMath*, https://epoch.ai/frontiermath (last visited Mar. 1, 2025).
[4] EPOCH AI, *FrontierMath*, https://epoch.ai/frontiermath (last visited Mar. 1, 2025).
[5] EPOCH AI, *FrontierMath*, https://epoch.ai/frontiermath (last visited Mar. 1, 2025).
[6] EPOCH AI, *FrontierMath*, https://epoch.ai/frontiermath (last visited Mar. 1, 2025), https://epoch.ai/frontiermath.
[7] *See* Maria Deutscher, *OpenAI details o3 reasoning model with record-breaking benchmark scores*, SILICONANGLE (Dec. 20, 2024) https://siliconangle.com/2024/12/20/openai-details-o3-reasoning-model-record-breaking-benchmark-scores/.
[8] *See* Maria Deutscher, *OpenAI details o3 reasoning model with record-breaking benchmark scores*, SILICONANGLE (Dec. 20, 2024) https://siliconangle.com/2024/12/20/openai-details-o3-reasoning-model-record-breaking-benchmark-scores/.
[9] *See* OPENAI, *Our Charter*, https://openai.com/charter/ (last visited Mar. 24, 2025) (stating that their intention is to develop AGI which could perform any economically valuable task).
[10] See e.g. "*Godfather of Artificial Intelligence" Weighs in on the Past and Potential of AI*, CBS NEWS (Mar. 25, 2023), https://www.cbsnews.com (last visited Oct. 7, 2024); *How Rogue AIs May Arise*, Yoshua Bengio (May 26, 2023), https://yoshuabengio.org (last visited Oct. 7, 2024); Alan Turing, *Intelligent Machinery, a Heretical Theory* (lecture, 51 Society, Manchester, 1951), in THE TURING DIGITAL ARCHIVE, https://www.turingarchive.ac.uk (last visited Oct. 7, 2024); Simon Parkin, *Science Fiction No More? Channel 4's Humans and Our Rogue AI Obsessions*, THE GUARDIAN (June 14, 2015), https://www.theguardian.com /tv-and-radio/2015/jun/14/science-fiction-no-more-humans-tv-artificial-intelligence (last visited Oct. 7, 2024); Sarah Jackson, *The CEO of the Company Behind AI Chatbot ChatGPT Says the Worst-Case Scenario for Artificial Intelligence is 'Lights Out for All of Us'*, BUS. INSIDER (Apr. 10, 2023), https://www.businessinsider.com /chatgpt-openai-ceo-worst-case-ai-lights-out-for-all-2023-1 (last visited Oct. 7, 2024).

This moment matters not only for mathematics, but for public institutions. If AI systems can now generate reasoning that rivals or exceeds that of domain experts, legal and regulatory frameworks built on assumptions of human comprehension, responsibility, and predictability begin to break down. Foundation models—general-purpose systems trained at scale and adapted across diverse applications—amplify this institutional challenge. As Bommasani et al. observe, these models blur the boundaries between capabilities, raise systemic risks, and outpace current governance mechanisms.[13]

Yet for all the attention paid to technical safeguards and governance frameworks, the conversation around AI policy has largely neglected a deeper structural challenge: how to manage the financial fallout from failure. If advanced AI systems behave in ways that are misaligned with human interests, the result may not be regulatory noncompliance but widespread economic damage or catastrophic harm. These are not hypothetical risks. As AI systems become more powerful and autonomous, the consequences of misalignment may spread faster than our ability to assign responsibility. The question, then, is not only how to control these systems, but how to anticipate, absorb, and respond to the damage when control fails.

In domains where uncertainty, liability, and harm meet, insurers allocate risk.[14] Yet private insurers remain hesitant to cover AI—opaque risks, uncertain outcomes.[15] Without a credible financial framework for catastrophic loss, the insurance market remains underdeveloped. A federal reinsurance program—used in contexts such as nuclear energy,[16] agriculture,[17] healthcare,[18] terrorism,[19] and natural disaster[20]—could fill the gap.

---

[11] *See* STUART J. RUSSELL & PETER NORVIG, ARTIFICIAL INTELLIGENCE: A MODERN APPROACH 1036 (4th ed. 2021).

[12] See *Our Approach to Alignment Research*, OPENAI (Oct. 24, 2022), https://openai.com/research/our-approach-to-alignment-research. (quoting OpenAI's admission of alignment uncertainty)

[13] *See* Rishi Bommasani et al., *On the Opportunities and Risks of Foundation Models* 4–5, 56–57 STANFORD INST. FOR HUMAN-CENTERED. ARTIFICIAL INTELLIGENCE (Aug. 2021), https://arxiv.org/abs/2108.07258.

[14] *See* Anat Lior, *Insuring AI: The Role of Insurance in Artificial Intelligence Regulation*, 35 HARV. J.L. & TECH. 468, 485–87 (2022).

[15] *See* Anat Lior, *Insuring AI: The Role of Insurance in Artificial Intelligence Regulation*, 35 HARV. J.L. & TECH. 468, 490–93 (2022).

[16] *See, e.g.* Price-Anderson Act, Pub. L. No. 85-256, 71 Stat. 576 (1957) (codified as amended at 42 U.S.C. § 2210 (2023).

[17] Federal Crop Insurance Act, 7 U.S.C. §§ 1501–1524 (2023).

[18] Patient Protection and Affordable Care Act, Pub. L. No. 111-148, § 1341, 124 Stat. 119, 208–11 (2010).

[19] Terrorism Risk Insurance Act of 2002, 15 U.S.C. §§ 6701–6711 (2023).

[20] National Flood Insurance Act of 1968, Pub. L. No. 90-448, 82 Stat. 572 (codified as amended at 42 U.S.C. §§ 4001–4128).

In high-risk industries, insurers shape conduct by pricing risk into coverage.[21] They exclude unsafe practices, refine standards, and reward compliance.[22] The same logic applies for frontier AI. A robust insurance market, secured by federal reinsurance, would complement direct regulation by conditioning coverage on transparency, monitoring, and adherence to safety norms.[23] The insurance industry already plays this role in medicine, aviation, and cybersecurity.[24] The structure is already in place.

Federal reinsurance enables markets to function where risk is uninsurable. Floods,[25] crop failures,[26] terrorism[27]—each needed public intervention to absorb tail risk and encourage private participation. Frontier AI is no different. Given the scale of unknown risks, a purely private insurance market will not form without public support.[28]

Critics warn that regulators may (1) delay technical advancement and (2) exceed their institutional understanding.[29] A federal reinsurance program meets both concerns. Insurers have skin in the game.[30] Their methods—structured, adaptive, accountable—create decentralized pressure toward safety.[31] Insurance firms function as learning institutions, assessing risk and identifying new vulnerabilities. Already, foundational research on systemic risk has been coauthored by reinsurers and those working in AI safety.[32]

In this context, insurance contracts are a form of soft regulation with teeth.[33] They discourage dangerous AI practices not by banning them, but by *making them*

---

[21] *See* Anat Lior, *Insuring AI: The Role of Insurance in Artificial Intelligence Regulation*, 35 HARV. J.L. & TECH. 468, 518 (2022); Kenneth S. Abraham & Catherine M. Sharkey, *The Glaring Gap in Tort Theory*, 133 YALE L.J. 1, 133 (2024).
[22] *See* Anat Lior, *Insuring AI: The Role of Insurance in Artificial Intelligence Regulation*, 35 HARV. J.L. & TECH. 468, 518 (2022).
[23] *See* Anat Lior, *Insuring AI: The Role of Insurance in Artificial Intelligence Regulation*, 35 HARV. J.L. & TECH. 468, 518 (2022).
[24] *See* Kenneth S. Abraham & Catherine M. Sharkey, *The Glaring Gap in Tort Theory*, 133 YALE L.J. 1, 133 (2024).
[25] *See* National Flood Insurance Act of 1968, Pub. L. No. 90-448, 82 Stat. 572 (codified as amended at 42 U.S.C. §§ 4001–4128).
[26] *See* Federal Crop Insurance Act, 7 U.S.C. §§ 1501–1524 (2023).
[27] *See* Terrorism Risk Insurance Act of 2002, 15 U.S.C. §§ 6701–6711 (2023).
[28] *See* Anat Lior, *Insuring AI: The Role of Insurance in Artificial Intelligence Regulation*, 35 HARV. J.L. & TECH. 468, 486, 502.
[29] *See* Fei-Fei Li, *Godmother of AI Warns SB 1047 AI Bill Restricts Innovation*, CAL. CHAMBER OF COM. (Aug. 7, 2024), https://advocacy.calchamber.com/2024/08/07/godmother-of-ai-warns-sb-1047-ai-bill-restricts-innovation/.
[30] *See* Anat Lior, *Insuring AI: The Role of Insurance in Artificial Intelligence Regulation*, 35 HARV. J.L. & TECH. 468, 511-13 (2022).
[31] *See* Anat Lior, *Insuring AI: The Role of Insurance in Artificial Intelligence Regulation*, 35 HARV. J.L. & TECH. 468, 511-13 (2022).
[32] *See e.g.* SYSTEMIC RISK OF MODELLING WORKING PARTY, DID YOUR MODEL TELL YOU ALL MODELS ARE WRONG? (Oxford Martin Sch. & Amlin, 2015), https://oms-www.files.svdcdn.com/production/downloads/academic/201511_Amlin_FHI_white_paper.pdf
[33] *See* Gary Marchant & Carlos Ignacio Gutierrez, *Soft Law 2.0: An Agile and Effective Governance Approach for Artificial Intelligence*, 24 MINN. J.L. SCI. & TECH. 375 (2023).

*expensive*.[34] A federal reinsurance program would not only stabilize the insurance market—it would promote both safety and innovation, creating a governance ecosystem that evolves with the field, rather than attempting to contain it.

**Federal reinsurance for advanced artificial intelligence offers a credible foundation for managing risk at scale.** Traditional legal tools—regulation, litigation, and voluntary guidelines—lack the institutional capacity to address deep uncertainty, widespread spillover effects, and low-probability but catastrophic harms. A public financial infrastructure distributes risk, incentivizes responsible development, and enables earlier detection of emerging threats. Precedent exists in nuclear energy, agriculture, healthcare, and finance, where federal reinsurance enabled markets to function despite underlying volatility. The same institutional logic applies to frontier AI.

Part I explains how general-purpose and frontier AI models work, and why they have become a major policy concern. Part II reviews extant legal responses, including regulatory efforts in the European Union and California, recent developments in tort law, and the role of voluntary frameworks. Part III identifies a deeper structural gap: existing institutions are not equipped to govern fast-moving, high-stakes risks of this kind. Part IV draws lessons from historical cases where federal reinsurance helped manage similarly complex and uncertain domains. Part V develops a concrete proposal: a three-tiered system combining required private insurance, a shared industry risk pool, and a federal reinsurance backstop. The Conclusion shows how this structure limits financial fallout and creates both the incentives and information needed to govern advanced AI in a serious, adaptive, and forward-looking way.

## I. BACKGROUND

### A. Technical Foundations of AI

In principle, artificial intelligence (AI) refers to anything that is both intelligent and made by humans.[35] In practice, the term denotes digital computers that simulate human cognition.[36] These systems perform tasks such as reasoning, problem solving, learning, and decision-making, tasks once required human intelligence.[37] While some research aims to mimic human capabilities, other efforts seek to build machines that exceed them.[38]

---

[34] *See* Gary Marchant & Carlos Ignacio Gutierrez, *Soft Law 2.0: An Agile and Effective Governance Approach for Artificial Intelligence*, 24 MINN. J.L. SCI. & TECH. 375 (2023).
[35] *See generally*, Alan M. Turing, *Computing Machinery and Intelligence*, 59 MIND 433, 433–60 (1950); STUART RUSSELL & PETER NORVIG, ARTIFICIAL INTELLIGENCE: A MODERN APPROACH 1–4 (4th ed. 2020). This definition is obviously not practicable because it arguably includes children.
[36] STUART RUSSELL & PETER NORVIG, ARTIFICIAL INTELLIGENCE: A MODERN APPROACH 1–4 (4th ed. 2020).
[37] STUART RUSSELL & PETER NORVIG, ARTIFICIAL INTELLIGENCE: A MODERN APPROACH 1–4 (4th ed. 2020).
[38] STUART RUSSELL & PETER NORVIG, ARTIFICIAL INTELLIGENCE: A MODERN APPROACH 1–4 (4th ed. 2020).

AI can be divided into three basic categories of capability: narrow AI, general AI, and superintelligent AI.[39] *Narrow* AI is also called weak AI; think of Siri answering your questions, Netflix recommending a show, or an algorithm sorting your emails.[40]

*General* AI, or strong AI, is a different beast. It is the next step, a machine that can think, reason, and adapt across a broad range of tasks, much like a human.[41] Imagine a program that can carefully explain how to fly a plane, pilot the plane by itself, and then write a compelling poem about the wonders of flight. Some experts believe we might get there in a few decades.[42] Others think true general AI is either impossible or a distant dream.[43]

Beyond that is *superintelligent* AI: machines that would not just match human intelligence but surpass it across every domain.[44] For now, it is pure speculation, but the implications are enormous. A superintelligent system could solve problems humanity has not even imagined or pose risks we are not ready to handle.[45]

*Machine learning* (ML) is a specific technique at the heart of modern AI.[46] Instead of following step-by-step instructions, machine learning algorithms learn from data by spotting patterns, making predictions, and improving over time.[47] There are different flavors. *Supervised learning* trains on labeled examples, like a student studying the answer key.[48] *Unsupervised learning* looks for hidden patterns in raw data, making sense of things without explicit guidance.[49] *Reinforcement learning* works through trial and error, adjusting its behavior based on rewards, much like training a dog with treats.[50]

Then there is *deep learning*, a powerful offshoot of machine learning that relies on what are called *multilayered neural networks*.[51] It is what makes facial recognition work, helps voice assistants understand speech, and allows AI to generate realistic images.[52]

---

[39] Shane Legg & Marcus Hutter, *A Collection of Definitions of Intelligence* in FRONTIERS IN ARTIFICIAL INTELLIGENCE AND APPLICATIONS (B. Goertzel & P. Wang eds., 2007).
[40] RAYMOND T. NIMMER ET AL., INFORMATION LAW § 1:16 (2024).
[41] John McCarthy, *What Is Artificial Intelligence?*, STANFORD UNIV. 2 (2007), http://www-formal.stanford.edu/jmc/whatisai/whatisai.html.
[42] Max Roser, *AI timelines: What do experts in artificial intelligence expect for the future?*, OUR WORLD IN DATA (Feb. 7, 2023) https://ourworldindata.org/ai-timelines.
[43] Max Roser, *AI timelines: What do experts in artificial intelligence expect for the future?*, OUR WORLD IN DATA (Feb. 7, 2023) https://ourworldindata.org/ai-timelines.
[44] RAYMOND T. NIMMER ET AL., INFORMATION LAW § 1:16 (2024).
[45] Ronald Bailey, *Will Superintelligent Machines Destroy Humanity?*, REASON, (Sept. 12, 2014) https://reason.com/2014/09/12/will-superintelligent-machines-destroy-h/.
[46] *See* RAYMOND T. NIMMER ET AL., INFORMATION LAW § 1:16 (2024).
[47] *See* RAYMOND T. NIMMER ET AL., INFORMATION LAW § 1:16 (2024).
[48] *See* RAYMOND T. NIMMER ET AL., INFORMATION LAW § 1:16 (2024).
[49] *See* RAYMOND T. NIMMER ET AL., INFORMATION LAW § 1:16 (2024).
[50] *See* RAYMOND T. NIMMER ET AL., INFORMATION LAW § 1:16 (2024).
[51] *See* RAYMOND T. NIMMER ET AL., INFORMATION LAW § 1:16 (2024).
[52] *See* RAYMOND T. NIMMER ET AL., INFORMATION LAW § 1:16 (2024).

Another key domain is *natural language processing* (NLP), which teaches machines to understand and produce human language.[53] That is how chatbots, translation tools, and voice assistants manage to sound so natural.[54] But even with all these advances, machine learning and deep learning are still forms of narrow AI.[55] They are impressive, but they do not think like humans do. They excel at specific tasks, but they do not truly understand what they are doing. For now, AI remains a powerful tool, but still far from the kind of intelligence that could rival a human being.[56]

**B. Current Applications of Artificial Intelligence**

AI is already changing industries in ways both big and small. In healthcare, it helps doctors diagnose diseases, personalize treatments[57] and speed up drug discovery.[58] In finance, AI spots fraud, powers algorithmic trading, and refines credit scoring, making decisions that once took hours happen in seconds.[59] Transportation is feeling the shift too, with self-driving cars learning to navigate city streets and AI predicting traffic accidents before they happen.[60] Meanwhile, the entertainment industry also runs on AI. Streaming services know what you will want to watch before you do,[61] and AI-powered tools can generate scripts, art, and even music.[62] Even in law, a world of dense paperwork and time-consuming research, AI speeds up document review and helps lawyers find relevant cases in minutes.[63] These are not just gimmicks. They are real shifts in how work gets done. And as AI continues to evolve, its roles in these fields will only grow.

AI is full of promise, but it also comes with serious challenges. Bias is a major problem. AI learns from the datasets it is given, and if those datasets contain bias, the system will pick it up and run with it, sometimes in ways that lead to unfair or even discriminatory decisions.[64] Privacy is another concern. Many AI systems thrive on personal data, raising questions about who controls that information and how it is being used.[65] Then there is the fear of job loss.[66] As AI gets better at automating tasks, entire industries could be disrupted, leaving workers wondering where they fit in.[67] And in high-stakes fields like healthcare and defense, the risks are even greater. When lives are on the line, AI needs to be not just smart, but predictable and reliable. The challenge is not just making AI more powerful, it is making sure we can trust it.

---

[53] *See* RAYMOND T. NIMMER ET AL., INFORMATION LAW § 1:16 (2024).
[54] *See* DANIEL JURAFSKY & JAMES H. MARTIN, SPEECH AND LANGUAGE PROCESSING (3d ed. 2025), https://web.stanford.edu/~jurafsky/slp3/.
[55] *See* RAYMOND T. NIMMER ET AL., INFORMATION LAW § 1:16 (2024).
[56] *See generally* Patrick Altmeyer et al., *Position Stop Making Unscientific AGI Performance Claims*, arXiv:2402.03962 [cs.AI] (2024), https://arxiv.org/abs/2402.03962.
[57] *See* Kevin B. Johnson et al., *Artificial Intelligence in Personalized Medicine: Current Trends and Future Perspectives*, PUBMED CENTRAL, https://pmc.ncbi.nlm.nih.gov/articles/PMC7877825 (Oct. 12, 2020).
[58] *See* Dolores R. Serrano, *The Role of Artificial Intelligence in Drug Discovery and Development*, PUBMED CENTRAL, https://pmc.ncbi.nlm.nih.gov/articles/PMC11510778 (Oct. 14, 2025).
[59] *What is artificial intelligence (AI) in finance?*, IBM (Dec. 8, 2023), https://www.ibm.com/think/topics/artificial-intelligence-finance.
[60] *A Blueprint for AV Safety: Waymo's Toolkit For Building a Credible Safety Case*, WAYMO (2023), https://waymo.com/safety/2023-safety-report.
[61] Xavier Amatriain & Justin Basilico, *Netflix Recommendations: Beyond the 5 Stars*, NETFLIX TECH BLOG (Apr. 6, 2012), https://netflixtechblog.com/netflix-recommendations-beyond-the-5-stars-part-1-55838468f429.
[62] See generally *About*, SUNO, https://suno.com/about (last visited Mar. 24, 2025)
[63] Harry Surden, *Machine Learning and Law*, 89 WASH. L. REV. 87, 99–104 (2014). https://digital-commons.law.uw.edu/wlr/vol89/iss1/5/

**C. The Debate Over AGI**

The prospect of artificial general intelligence (AGI) spurs debate among experts. Researchers at the cutting edge of machine intelligence wrestle with questions of how to design a safe AGI, yet critics argue that such efforts remain highly speculative.[68] They maintain that true AGI demands integrated reasoning, creativity, and common sense across a broad assortment of tasks—capabilities beyond current AI's reach.[69] Skeptics further posit that the realization of AGI may be decades away or might never happen at all, depending on how one defines "intelligence" and whether extant technical barriers can be overcome.[70]

---

[64] Joy Buolamwini & Timnit Gebru, *Gender Shades: Intersectional Accuracy Disparities in Commercial Gender Classification*, in PROCEEDINGS OF THE 1ST CONF. ON FAIRNESS, ACCOUNTABILITY AND TRANSPARENCY 77, 77–91 (S. A. Friedler & C. Wilson eds., 2018). https://proceedings.mlr.press/v81/buolamwini18a.html.

[65] *See* Lilian Mitrou, *Data Protection, Artificial Intelligence and Cognitive Services: Is the General Data Protection Regulation (GDPR) 'Artificial Intelligence-Proof'?*, 2 EUR. DATA PROT. L. REV. 20 (2018), https://www.researchgate.net/publication/344746896_Data_Protection_Artificial_Intelligence_and_Cognitive_Services_Is_the_General_Data_Protection_Regulation_GDPR_'Artificial_Intelligence-Proof'.

[66] *See* Carl Benedikt Frey & Michael A. Osborne, *The future of employment: How susceptible are jobs to computerization?*, in 114 TECHNOLOGICAL FORECASTING AND SOCIAL CHANGE 254 (2017), https://www.sciencedirect.com/science/article/abs/pii/S0040162516302244.

[67] *See* Carl Benedikt Frey & Michael A. Osborne, *The future of employment: How susceptible are jobs to computerization?*, in 114 TECHNOLOGICAL FORECASTING AND SOCIAL CHANGE 254 (2017), https://www.sciencedirect.com/science/article/abs/pii/S0040162516302244.

[68] *See* Parmy Olson, *Meta's AI Chief Yann Le Cun Says Fears about the Technology are Overblown*, WALL ST. J. (Feb. 15, 2024), https://www.wsj.com/tech/ai/yann-lecun-ai-meta-aa59e2f5; Kelvin Chan, *Global AI Report Warns of "Evidence Dilemma" in Addressing Potential Risks*, AP (Feb. 26, 2025), https://apnews.com/article/7b9db4ca69a89a4dd04e05a4294a3dfd; Henry Kautz, *The Curious Case of Commonsense Intelligence*, 151 DAEDALUS 139, 139–50 (2022), https://direct.mit.edu/daed/article/151/2/139/110627/The-Curious-Case-of-Commonsense-Intelligence.

[69] Parmy Olson, *Meta's AI Chief Yann Le Cun Says Fears about the Technology are Overblown*, WALL ST. J. (Feb. 15, 2024), https://www.wsj.com/tech/ai/yann-lecun-ai-meta-aa59e2f5; Kelvin Chan, *Global AI Report Warns of "Evidence Dilemma" in Addressing Potential Risks*, AP (Feb. 26, 2025), https://apnews.com/article/7b9db4ca69a89a4dd04e05a4294a3dfd; Henry Kautz, *The Curious Case of Commonsense Intelligence*, 151 DAEDALUS 139, 139–50 (2022), https://direct.mit.edu/daed/article/151/2/139/110627/The-Curious-Case-of-Commonsense-Intelligence.

[70] Parmy Olson, *Meta's AI Chief Yann Le Cun Says Fears about the Technology are Overblown*, WALL ST. J. (Feb. 15, 2024), https://www.wsj.com/tech/ai/yann-lecun-ai-meta-aa59e2f5; Kelvin Chan, *Global AI Report Warns of "Evidence Dilemma" in Addressing Potential Risks*, AP (Feb. 26, 2025), https://apnews.com/article/7b9db4ca69a89a4dd04e05a4294a3dfd; Henry Kautz, *The Curious Case of Commonsense Intelligence*, 151 DAEDALUS 139, 139–50 (2022).

Philosopher and mathematician Roger Penrose, for instance, contends that human consciousness eludes purely algorithmic explanation.[71] In *The Emperor's New Mind*, he invokes Gödel's Incompleteness Theorems to suggest that human beings can perceive truths that formal systems cannot prove, indicating that the mind exceeds computational confines.[72] Penrose further speculates that quantum processes in the brain may play a vital role in consciousness—a hypothesis that remains subject to ongoing scientific and philosophical scrutiny.[73]

Despite lingering doubts about AI's ultimate frontiers, major technology companies vigorously pursue more advanced and general AI. Apple, Microsoft, and Google (now Alphabet), among others, remain at the vanguard of research, leveraging immense resources to stake a claim in the race for ever-more capable systems.[74] Google's subsidiary, DeepMind, has produced two products: AlphaZero, which consistently bests all humans in chess, shogi, and Go, and AlphaFold which surpasses expert performance in predicting protein folding—to the chagrin of the entire biopharmaceutical R&D industry.[75] Facebook (now Meta) has introduced *CICERO*, an AI designed for the strategy game *Diplomacy*, which requires negotiation, deceit, and alliance-building.[76] The ability of such systems to perform at or above human levels in varied tasks underscores AI's accelerating progress toward broader forms of intelligence.

Artificial intelligence continues to evolve at an extraordinary pace, bringing profound transformations to multiple facets of society. While narrow AI dominates contemporary applications, research on more sophisticated systems nudges the field closer to general—if not superintelligent—forms of machine cognition. Yet as these capabilities expand, so do the attendant ethical, legal, and societal questions concerning safety, privacy, and the nature of intelligence itself. Navigating these challenges demands not only technical innovation but also robust interdisciplinary collaboration to ensure that AI's development proceeds responsibly and equitably.

---

[71] ROGER PENROSE, THE EMPEROR'S NEW MIND: CONCERNING COMPUTERS, MINDS AND THE LAWS OF PHYSICS 132-141 (1989)

[72] ROGER PENROSE, THE EMPEROR'S NEW MIND: CONCERNING COMPUTERS, MINDS AND THE LAWS OF PHYSICS 132–141 (1989)

[73] ROGER PENROSE, THE EMPEROR'S NEW MIND: CONCERNING COMPUTERS, MINDS AND THE LAWS OF PHYSICS 132 –141 (1989)

[74] Julia Kollewe, *Apple Cheers Trump with $500bn US Investment Plan; More Losses on Wall Street – As It Happened*, THE GUARDIAN (Feb. 24, 2025), https://www.theguardian.com/business/live/2025/feb/24/euro-hits-one-month-high-german-election-result-stock-markets-dax-bank-of-england-business-live-news; *Tech Giants to Spend $320 Billion on AI in 2025: Meta, Amazon, Alphabet, Microsoft Lead the Race—What About Apple, Tesla, and Nvidia?*, ECON. TIMES (Feb. 8, 2025), https://m.economictimes.com/news/international/us/tech-giants-to-spend-320-billion-on-ai-in-2025-meta-amazon-alphabet-microsoft-lead-the-race-what-about-apple-tesla-and-nvidia/articleshow/118068850.cms.

[75] David Silver et al., *Mastering the Game of Go with Deep Neural Networks and Tree Search*, 529 NATURE 484, 484–89 (2016); John Jumper et al., *Highly Accurate Protein Structure Prediction with AlphaFold*, 596 NATURE 583, 583–89 (2021).

[76] Noam Brown et al., *Human-Level Play in the Game of Diplomacy by Combining Language Models with Strategic Reasoning*, 378 SCIENCE 1067, 1067-74 (2022). Andrew Goff et al., *CICERO: An AI Agent That Negotiates, Persuades, and Cooperates with People*, META AI (Nov. 22, 2022), https://ai.meta.com/blog/cicero-ai-negotiates-persuades-and-cooperates/.

## II. THE LEGAL LANDSCAPE OF AI

As AI systems grow more sophisticated, they present new challenges for regulation, liability, and enforcement. Governments worldwide are grappling with how to regulate AI effectively without stifling innovation. At the same time, tort law, traditionally designed for human actors, must now account for autonomous systems that make decisions without direct human input. Alongside these formal legal mechanisms, *soft law*, nonbinding principles and guidelines, is emerging as a flexible tool for shaping AI governance. Together, these three areas form the foundation of how society seeks to balance the promise of AI with the need for oversight and accountability.

### A. Regulatory Approaches

Regulating AI is a delicate task. Unlike traditional technologies, AI evolves, learns, and adapts, making regulatory frameworks difficult to apply. Policymakers must strike a balance between fostering innovation and preventing harm. Different jurisdictions have taken different approaches. The European Union has opted for comprehensive, preemptive regulation, while the United States has favored a more fragmented, sector specific strategy. These differing approaches highlight the complexity of AI governance and the competing interests at play.

*1. The European Union*

The European Union (EU) has taken a proactive stance on AI governance. The Artificial Intelligence Act (AI Act), which came into force on August 1, 2024, establishes a harmonized legal framework across Member States.[77] The AI Act categorizes AI systems by risk level, unacceptable, high, limited, and minimal, each with corresponding regulatory requirements.[78] Systems deemed "unacceptable," such as those manipulate human behavior through subliminal techniques, are outright banned.[79] "High-risk" AIs, including those in critical infrastructure and education, must meet stringent transparency and oversight standards before deployment.[80]

---

[77] *See generally,* Krystyna Marcinek et al., *Risk-Based AI Regulation: A Primer on the Artificial Intelligence Act of the European Union*, RAND (Nov. 20, 2024) https://www.rand.org/pubs/research_reports/RRA3243-3.html#fnb7.

[78] Regulation (EU) 2024/1689 of the European Parliament and of the Council of 13 June 2024, Laying Down Harmonised Rules on Artificial Intelligence (Artificial Intelligence Act) and Amending Certain Union Legislative Acts, 2024 O.J. (L 379) 1, arts. 5, annex III; *Europe's AI Act: The New Rules*, EUR. COMM'N (Feb. 19, 2025), https://digital-strategy.ec.europa.eu/en/policies/regulatory-framework-ai.

[79] Regulation (EU) 2024/1689 of the European Parliament and of the Council of 13 June 2024 Laying Down Harmonised Rules on Artificial Intelligence (Artificial Intelligence Act) and Amending Certain Union Legislative Acts, 2024 O.J. (L 379) 1, art. 5.

[80] Regulation (EU) 2024/1689 of the European Parliament and of the Council of 13 June 2024, Laying Down Harmonised Rules on Artificial Intelligence (Artificial Intelligence Act) and Amending Certain Union Legislative Acts, 2024 O.J. (L 379) 1, art. 5(1)(h), 5(2)–(5).

To avoid stifling innovation, the AI Act contains provisions to ease regulatory burdens on small and medium sized enterprises.[81] Additional initiatives, such as the AI Innovation Package and the Coordinated Plan on AI, support AI development while enforcing compliance with ethical and safety standards.[82] By establishing clear obligations and enforcement mechanisms, the EU seeks to set the global standard on AI governance.[83]

2. *The United States*

In contrast to the EU's centralized approach, the United States has adopted a more decentralized, patchwork strategy. Federal initiatives, state legislation, and international collaborations each play a role in shaping their approach to AI governance. The *National Artificial Intelligence Initiative Act* of 2020 laid the foundation for coordinated AI research and development across federal agencies.[84] In October 2023, President Biden issued Executive Order 14110, emphasizing AI safety, competition, and civil rights protections.[85] A year later, the administration issued a National Security Memorandum outlining AI's role in defense and intelligence operations.[86] Additionally, the Department of Commerce established the United States AI Safety Institute within the National Institute of Standards and Technology (NIST) to create guidelines and best practices for evaluating and mitigating AI risks.[87]

Internationally, the U.S. has promoted responsible AI use through initiatives such as the organization of American States' AI Policy Framework and the State

---

[81] Regulation (EU) 2024/1689 of the European Parliament and of the Council of 13 June 2024, Laying Down Harmonised Rules on Artificial Intelligence (Artificial Intelligence Act) and Amending Certain Union Legislative Acts, 2024 O.J. (L 379) 1, art. 6, annex I.

[82] Commission Communication on Fostering a European Approach to Artificial Intelligence, COM (2021) 205 final (Apr. 21, 2021); European Commission, *Coordinated Plan on Artificial Intelligence (2021 Review)*, COM (2021) 205 final (Apr. 21, 2021).

[83] Commission Regulation 2024/1689 of the European Parliament and of the Council Laying Down Harmonised Rules on Artificial Intelligence (Artificial Intelligence Act) and Amending Certain Union Legislative Acts, 2024 O.J. (L 1689) 1.

[84] National Artificial Intelligence Initiative Act of 2020, Pub. L. No. 116-283, div. E, § 5001, 134 Stat. 4523 (2021).

[85] Exec. Order No. 14,110, 88 Fed. Reg. 75,191 (Oct. 30, 2023).

[86] *Memorandum on Advancing the United States Leadership in Artificial Intelligence: Harnessing Artificial Intelligence to Fulfill National Security Objectives and Fostering the Safety and Security of Artificial Intelligence*, 2024 DAILY COMP. PRES. DOC. (Oct. 24, 2024), https://www.whitehouse.gov/briefing-room/presidential-actions/2024/10/24/memorandum-on-advancing-the-united-states-leadership-in-artificial-intelligence-harnessing-artificial-intelligence-to-fulfill-national-security-objectives-and-fostering-the-safety-security; *Fact Sheet, Biden-Harris Administration Outlines Coordinated Approach to Harness Power of AI for U.S. National Security*, WHITE HOUSE (Oct. 24, 2024), https://www.whitehouse.gov/briefing-room/statements-releases/2024/10/24/fact-sheet-biden-harris-administration-outlines-coordinated-approach-to-harness-power-of-ai-for-u-s-national-security.

[87] *Fact Sheet, U.S. Department of Commerce & U.S. Department of State Launch International AI Safety Initiative*, U.S. DEP'T OF STATE (Nov. 20, 2024), https://www.commerce.gov/news/fact-sheets/2024/11/fact-sheet-us-department-commerce-us-department-state-launch-international

Department's Bureau for Cyberspace and Digital Policy.[88] While the U.S. regulatory landscape remains fragmented, these efforts signal a growing recognition of AI's risks and the need for oversight.

State governments have also taken the lead. Idaho's 2024 House Bill 382, addressed AI's role in crimes against children, reflecting broader efforts by states to regulate AI's societal impact.[89] Moreover, the National Conference of State Legislatures has documented various AI-related legislative efforts across different states, reflecting a growing recognition of AI's impact on society.[90]

*3. California Senate Bill 1047*

California, often a leader in tech policy, attempted to introduce a comprehensive AI regulatory framework through Senate Bill 1047 (SB 1047), the *Safe and Secure Innovation for Frontier Artificial Intelligence Models Act*.[91] The bill sought to enhance transparency, and hold developers accountable for their technologies' societal impacts.[92] The aspect of the bill that was noteworthy for the argument in this Note is that it focused not on all AI, but on a specific subset of high-risk systems, introducing the term "*covered model*" to define the types of AI subject to enhanced oversight. Under the bill, a covered model included any generative AI system trained using computer power (compute) exceeding $10^{26}$ FLOPS, or one that the developer had reason to believe could independently perform tasks that pose a sever risk to public safety, such as designing biological or chemical weapons.[93] However, Governor Gavin Newsom vetoed SB 1047 on September 29, 2024, arguing that the bill lacked flexibility to keep pace with AI's rapid evolution.[94] Instead, he announced alternative initiatives to safeguard Californians from AI-related risks.[95]

---

[88] *U.S. Mission to the Organization of American States Launches New Initiative on Artificial Intelligence*, U.S. DEP'T OF STATE (2024), https://2021-2025.state.gov/u-s-mission-to-the-organization-of-american-states-launches-new-initiative-on-artificial-intelligence/; *Bureau of Cyberspace and Digital Policy*, U.S. DEP'T OF STATE, https://www.state.gov/bureaus-offices/deputy-secretary-of-state/bureau-of-cyberspace-and-digital-policy.
[89] H.B. 382, 67th Leg., 2d Reg. Sess. (Idaho 2024), https://legislature.idaho.gov/session-info/2024/legislation/H0382.
[90] *Artificial Intelligence 2024 Legislation*, NAT'L CONF. OF STATE LEGISLATURES (Sept. 09, 2024), https://www.ncsl.org/technology-and-communication/artificial-intelligence-2024-legislation.
[91] S.B. 1047, 2023–2024 Leg., Reg. Sess. (Cal. 2024) (as vetoed by Governor, Sept. 29, 2024).
[92] S.B. 1047, 2023–2024 Leg., Reg. Sess. (Cal. 2024) (as vetoed by Governor, Sept. 29, 2024).
[93] S.B. 1047, 2023–2024 Leg., Reg. Sess. § 22601(e) (Cal. 2024) (defining "covered model" as a model trained using computational resources exceeding $10^{26}$ integer or floating-point operations or capable of autonomously performing tasks posing severe risk).
[94] Gavin Newsom, Veto Message for S.B. 1047 (Sept. 29, 2024), https://www.gov.ca.gov/wp-content/uploads/2024/09/SB-1047-Veto-Message.pdf.
[95] Governor Newsom Announces New Initiatives to Advance Safe and Responsible AI Development, Office of Governor Gavin Newsom (Sept. 28, 2024), https://www.gov.ca.gov/2024/09/28/governor-newsom-announces-new-initiatives-to-advance-safe-and-responsible-ai-development/.

Despite the veto, California remains at the forefront of AI regulation. Assembly Bill 2013 (AB 2013), effective January 1, 2026, mandates disclosure of training data used in generative AI systems.[96] Additional laws restrict AI's role in mental health services, preventing AI systems from impersonating human therapists.[97] While SB 1047 did not become law, California's regulatory efforts illustrate the state's desire for AI oversight. Later, this Note will use the definition of a "covered model" as a starting point for designed target, risk-based governance mechanism. As Governor Newsom rejected the bill for its lack of flexibility. This Note will argue for an alternative scheme of governance for these "covered models" that can keep pace with rapid technological development.

**B. Tort Law: Existing Doctrines & Emerging Challenges**

Tort law was built for human actors.[98] When someone causes harm through negligence or intent, the law holds them accountable.[99] What happens when an AI system causes harm? Who is responsible? The developer? The manufacturer? Or the AI system? The questions are at the heart of AI and tort law.[100]

Two recent cases—*Cruz v. Talmadge*[101] and *Nilsson v. General Motors, LLC*[102]—mark the beginning of a shift in product liability law.[103] They raise questions that courts have never had to answer before: When AI makes a mistake, who takes the blame? What does it mean for a machine to be "negligent"? Can a product itself be held liable? And if so, who—if anyone—pays?

The accident in *Cruz v. Talmadge* was as tragic as it was avoidable.[104] A bus, following the guidance of two GPS devices, drove straight into an overpass.[105] Passengers were injured.[106] Some were killed.[107] And their families wanted to know: Who was responsible? The bus driver had done what drivers always do—followed the GPS.[108] The AI-powered navigation system had all the data it needed to prevent the accident.[109] It knew the clearance.[110] It knew the risk. But it did not warn the driver. It didn't reroute the bus. And so the plaintiffs argued that this was not just a mistake—it was a defect.[111]

---

[96] A.B. 2013, 2023–2024 Leg., Reg. Sess. (Cal. 2024), https://leginfo.legislature.ca.gov/faces/billTextClient.xhtml?bill_id=202320240AB2013.
[97] *Assemblymember Mia Bonta Introduces Legislation to Prevent AI Systems from Impersonating Human Therapists*, Office of Assemblymember Mia Bonta (Feb. 9, 2025), https://a18.asmdc.org/press-releases/20250210-assemblymember-mia-bonta-introduces-legislation-prevent-ai-systems.
[98] *See* Cong. Rsch. Serv., Introduction to Tort Law, IF11291, at 1 (2023).
[99] *See* Cong. Rsch. Serv., Introduction to Tort Law, IF11291, at 1 (2023).
[100] *See* Rebecca Crootof, *The Internet of Torts*, 69 Duke L.J. 583, 585–88 (2019); *See also* Gregory Smith et al., *Liability for Harms from AI Systems: The Application of U.S. Tort Law and Liability to Harms from Artificial Intelligence Systems*, RAND 2024, https://www.rand.org/pubs/research_reports/RRA3243-4.html.
[101] Cruz v. Talmadge, 244 F. Supp. 3d 231 (D. Mass. 2017).
[102] Nilsson v. General Motors LLC, No. 3:18-cv-00471 (N.D. Cal. filed Jan. 22, 2018).
[103] Cruz v. Talmadge, 244 F. Supp. 3d 231, 233 (D. Mass. 2017).
[104] Cruz v. Talmadge, 244 F. Supp. 3d 231, 233 (D. Mass. 2017).
[105] Cruz v. Talmadge, 244 F. Supp. 3d 231, 233 (D. Mass. 2017).
[106] Cruz v. Talmadge, 244 F. Supp. 3d 231, 233 (D. Mass. 2017).
[107] Cruz v. Talmadge, 244 F. Supp. 3d 231, 233 (D. Mass. 2017).
[108] Cruz v. Talmadge, 244 F. Supp. 3d 231, 233 (D. Mass. 2017).

Their case raised a fundamental question: when an AI-powered product leads someone into danger, is the manufacturer liable for what happens next? And more than that—what does "reasonable care" mean when no human made the decision? Courts have long asked whether a person acted as a "reasonable driver" or a "reasonable doctor" or a "reasonable manufacturer."[112] But when the decision belongs to a machine, how do you determine what a reasonable machine should have done?[113]

In *Nilsson v. General Motors, LLC*, a motorcyclist was riding on the highway when an autonomous vehicle—a self-driving Chevrolet Bolt—swerved into his lane.[114] He crashed. He was injured.[115] And when he took General Motors to court, he made a striking claim: this was not driver error. This was the car's fault.[116] There was a backup driver behind the wheel, but he was not operating the car at the moment of impact.[117] The AI was driving.[118] And if a human driver can be negligent—if a human driver can be sued for failing to use reasonable care—why should not the same be true for an AI?[119] General Motors did not fight the premise.[120] In fact, it admitted that the Bolt was required to meet the same standard of care as a human driver.[121] That admission was a turning point. But it left behind an even bigger question: if an AI-powered vehicle is negligent, who pays the price?

Because AI does not belong to a single person. The car had an owner. The software had engineers. The company had designers, executives, and shareholders. If an autonomous vehicle makes a bad decision, who should be responsible? The manufacturer? The owner? The company that designed the AI? The programmer who wrote the faulty line of code? The answer is not obvious. And as AI grows more autonomous, it will only become harder to find.[122] These cases show that the legal system is at the start of a transformation. AI is no longer just a tool—it is a decision-maker. It is guiding vehicles, choosing routes, determining risk. And when AI makes a bad decision, courts must answer three urgent questions:

1. *What does reasonable care mean for a machine?* Courts have long measured human behavior against what a reasonable person would do.[123] But how do you judge a machine's choices? Some scholars suggest looking to custom, practice, or outcome.[124] Others point out that AI models lack intent, emotion, or experience—qualities necessary to inform a negligence analysis.[125]
2. *How do you define foreseeability for AI?* Humans make mistakes. But AI operates on a massive number of data, with predictive capabilities far beyond a person.[126] If an AI-driven product causes harm, was that harm foreseeable? And if so, who should have foreseen it?
3. *If AI is liable, who pays?* A product is not a person. It cannot be sued, fined, or held accountable.[127] But if a self-driving car, a surgical robot, or a financial algorithm causes harm, courts must determine whether liability falls on the manufacturer, the software developer, the owner, or someone else entirely.[128]

---

[109] Cruz v. Talmadge, 244 F. Supp. 3d 231, 233 (D. Mass. 2017).
[110] Cruz v. Talmadge, 244 F. Supp. 3d 231, 233 (D. Mass. 2017).
[111] Cruz v. Talmadge, 244 F. Supp. 3d 231, 233 (D. Mass. 2017).
[112] RESTATEMENT (THIRD) OF TORTS: LIAB. FOR PHYS. & EMOT. HARM § 3 (AM. L. INST. 2010).
[113] *See* Mark Geistfeld, *Strict Products Liability 2.0*, 14 J. TORT L. 403, 419–20 (2021).
[114] Nilsson v. General Motors LLC, No. 3:18-cv-00471 (N.D. Cal. filed Jan. 22, 2018).
[115] Nilsson v. General Motors LLC, No. 3:18-cv-00471 (N.D. Cal. filed Jan. 22, 2018).
[116] Nilsson v. General Motors LLC, No. 3:18-cv-00471 (N.D. Cal. filed Jan. 22, 2018).
[117] Nilsson v. General Motors LLC, No. 3:18-cv-00471 (N.D. Cal. filed Jan. 22, 2018).
[118] Nilsson v. General Motors LLC, No. 3:18-cv-00471 (N.D. Cal. filed Jan. 22, 2018).
[119] Mark Geistfeld, *A Roadmap for Autonomous Vehicles*, 105 CALIF. L. REV. 1611, 1615–16 (2017).
[120] Nilsson v. General Motors LLC, No. 3:18-cv-00471 (N.D. Cal. filed Jan. 22, 2018).
[121] Nilsson v. General Motors LLC, No. 3:18-cv-00471 (N.D. Cal. filed Jan. 22, 2018).
[122] Nilsson v. General Motors LLC, No. 3:18-cv-00471 (N.D. Cal. filed Jan. 22, 2018); Kenneth S. Abraham & Catherine M. Sharkey, *The Glaring Gap in Tort Theory*, 133 YALE L.J. 2165, 2172–74 (2024).

Proposed solutions include algorithmic accountability, which holds developers liable for flawed AI decision making, and enterprise liability, which places responsibility on companies profiting from AI.[129] Insurance may also play a role, with specialized AI insurance pools spreading risk across industries.[130] As courts and legislatures confront these issues, new legal precedents will shape the evolving intersection of AI and tort law.

**C. Soft Law & Voluntary Governance**

Regulation is not the only way to govern AI. Soft law—nonbinding guidelines, ethical frameworks, and industry standards—often fills the gaps where formal laws lag behind.[131] Unlike statutes and regulations, soft law can adapt quickly to technological changes, providing a flexible approach to AI oversight.[132] Scholars have documented an explosion of such instruments in recent years, identifying recurring themes like transparency, fairness, accountability, and human oversight across dozens of frameworks worldwide.[133]

---

[123] *See* Mark P. Gergen, *The Jury's Role in Deciding Normative Issues in the American Common Law*, 68 FORDHAM L. REV. 407, 425 (1999).
[124] *See* The T.J. Hooper v. N. Barge Corp., 60 F.2d 737, 740 (2d Cir. 1932).
[125] *See* Chan et al., *Harms from Increasingly Agentic Systems*, ARXIV (2023), https://arxiv.org/abs/2302.10329; Kenneth S. Abraham, *Custom, Noncustomary Practice, and Negligence*, 109 COLUM. L. REV. 1784 (2009).
[126] *See* Chan et al., *Harms from Increasingly Agentic Systems*, ARXIV (2023), https://arxiv.org/abs/2302.10329.
[127] *See* CONG. RSCH. SERV., INTRODUCTION TO TORT LAW, IF11291, at 1 (2023)..
[128] Mariano-Florentino Cuéllar, *A Common Law for the Age of Artificial Intelligence* 119 COLUM. L. REV. 1773, 1781–82 (2019).
[129] Catherine M. Sharkey, *Public Nuisance as Modern Business Tort*, 70 DEPAUL L. REV. 431, 432–33(2020).
[130] *See generally* Kenneth S. Abraham & Daniel Schwarcz, *Courting Disaster*, 27 CONN. INS. L.J. 407 (2021); *See also* Samuel R. Gross & Kent D. Syverud, *Don't Try*, 44 UCLA L. REV. 1, 5–7 (1996)..
[131] *See* Gary Marchant & Carlos Ignacio Gutierrez, *Soft Law 2.0: An Agile and Effective Governance Approach for Artificial Intelligence*, 24 MINN. J.L. SCI. & TECH. 375, 376 (2023).

Governments, international bodies, and industry groups use soft law to establish best practices without imposing legal mandates.[134] The universal Declaration of Human Rights, for example, set norms before being incorporated into binding treaties.[135] Similarly, AI soft law includes guidelines from organizations like the OECD,[136] the European Commission's Ethical Guidelines for Trustworthy AI,[137] and corporate AI ethics statements.[138] Mapping studies by researchers at Harvard's Berkman Klein Center[139] and ETH Zurich[140] reveal substantial international convergence around these principles, even as enforcement mechanisms remain absent. Soft law's strength lies in its ability to shape norms and influence behavior without legal coercion.[141] However, its weakness is its lack of enforceability.[142] Still, soft law often serves as a steppingstone to formal regulation.[143] It provides an indirect, adaptive mechanism for encouraging safety practices, shaping norms,

---

[132] *See* Gary Marchant & Carlos Ignacio Gutierrez, *Soft Law 2.0: An Agile and Effective Governance Approach for Artificial Intelligence*, 24 MINN. J.L. SCI. & TECH. 375, 377–378 (2023).

[133] Anna Jobin, Marcello Ienca & Effy Vayena, *The Global Landscape of AI Ethics Guidelines*, 1 NAT. MACH. INTELL. 389 (2019); *AI Ethics Guidelines Global Inventory* ALGORITHM WATCH (2019) https://algorithmwatch.org/en/ai-ethics-guidelines-global-inventory/.

[134] Gary Marchant & Carlos Ignacio Gutierrez, *Soft Law 2.0: An Agile and Effective Governance Approach for Artificial Intelligence*, 24 MINN. J.L. SCI. & TECH. 375, 384 (2023).

[135] G.A. Res. 217 (III) A, Universal Declaration of Human Rights, U.N. Doc. A/RES/217(III) (Dec. 10, 1948); https://www.un.org/sites/un2.un.org/files/2021/03/udhr.pdf

[136] RECOMMENDATION OF THE COUNCIL ON ARTIFICIAL INTELLIGENCE, OECD (May 22, 2019) https://www.oecd.org/en/topics/artificial-intelligence.html.

[137] HIGH-LEVEL EXPERT GROUP ON ARTIFICIAL INTELLIGENCE, ETHICAL GUIDELINES FOR TRUSTWORTHY AI, EUROPEAN COMMISSION (Apr. 8, 2019) https://www.europarl.europa.eu/cmsdata/196377/AI%20HLEG_Ethics%20Guidelines%20for%20Trustworthy%20AI.pdf.

[138] *Responsible AI Principles,* MICROSOFT (2020) https://www.microsoft.com/en-us/ai/responsible-ai?ef_id=_k_b60ce3d9a01f17f83eaf84386dcddad3_k_&OCID=AIDcmm1o1fzy5i_SEM__k_b60ce3d9a01f17f83eaf84386dcddad3_k_&msclkid=b60ce3d9a01f17f83eaf84386dcddad3; Sundar Pichai, *AI at Google: our principles,* (Jun. 7, 2018) https://blog.google/technology/ai/ai-principles/; *Digital Ethics Guidelines on AI,* DEUTSCHE TELEKOM (2018) https://www.telekom.com/en/company/digital-responsibility/digital-ethics-deutsche-telekoms-ai-guideline

[139] Jessica Fjeld et al., *Principled Artificial Intelligence: Mapping Consensus in Ethical and Rights-Based Approaches to Principles for AI*, BERKMAN KLEIN CTR. RSCH. PUB. No. 2020-1 (2020).

[140] Anna Jobin, Marcello Ienca & Effy Vayena, *The Global Landscape of AI Ethics Guidelines*, 1 NAT. MACH. INTELL. 389 (2019)

[141] *See* Gary Marchant & Carlos Ignacio Gutierrez, *Soft Law 2.0: An Agile and Effective Governance Approach for Artificial Intelligence*, 24 MINN. J.L. SCI. & TECH. 375, 396 (2023).

[142] *See* Gary Marchant & Carlos Ignacio Gutierrez, *Soft Law 2.0: An Agile and Effective Governance Approach for Artificial Intelligence*, 24 MINN. J.L. SCI. & TECH. 375, 399 (2023).

[143] Jessica Fjeld et al., *Principled Artificial Intelligence: Mapping Consensus in Ethical and Rights-Based Approaches to Principles for AI*, BERKMAN KLEIN CTR. RSCH. PUB. No. 2020-1 (2020); Gary Marchant, Logan Tournas & Carlos I. Gutierrez, *Governing Emerging Technologies Through Soft Law: Lessons for Artificial Intelligence*, 61 JURIMETRICS 1, (2021).

and guiding institutional responses.[144] In this way, soft law and market-based strategies like reinsurance can work in tandem to govern frontier AI development, offering scalable alternatives to direct regulatory intervention.[145]

### III. THE GOVERNANCE GAP

Efforts to regulate artificial intelligence face a structural asymmetry that has long challenged administrative law: legislation moves slowly, but technology evolves at exponential speed.[146] By the time statutory frameworks are drafted, debated, and enacted, the systems they were meant to govern have often already shifted.[147] Policymakers face a difficult design problem—crafting rules that are both future-proof and capable of constraining real risks in the present.[148] Yet even this challenge understates the problem. Many modern AI systems, especially large foundation models, are epistemically opaque.[149] Their inner workings are difficult to interpret, even for their developers.[150] This opacity complicates the task of regulatory design and undermines enforcement, making it hard to establish both *ex ante* constraints and *ex post* accountability.[151]

Tort law, the common law's traditional mechanism for assigning liability,[152] is similarly strained.[153] When a self-driving car crashes, or a foundation model produces hazardous content, the causal chain is often too complex to trace using traditional fault-based doctrines.[154] Plaintiffs struggle to establish breach, foreseeability, or proximate cause when harm emerges from probabilistic systems trained on vast and dynamic datasets.[155] In many cases, developers may themselves lack a clear explanation of how their models arrived at a harmful output.[156] If liability becomes functionally unprovable, victims remain uncompensated and deterrence fails. Conversely, if liability is imposed too broadly or unpredictably, innovation may be chilled.[157]

---

[144] Gary Marchant, Logan Tournas & Carlos I. Gutierrez, *Governing Emerging Technologies Through Soft Law: Lessons for Artificial Intelligence*, 61 JURIMETRICS 1, (2021).

[145] Gary Marchant & Carlos Ignacio Gutierrez, *Soft Law 2.0: An Agile and Effective Governance Approach for Artificial Intelligence*, 24 MINN. J.L. SCI. & TECH. 375, 424 (2023); Gary E. Marchant & Braden Allenby, *Soft Law: New Tools for Governing Emerging Technologies*, 73 BULL. ATOMIC SCI. 1 (2017).

[146] *See* Rishi Bommasani et al., *On the Opportunities and Risks of Foundation Models*, arXiv:2108.07258 (July 12, 2022).

[147] *See* Rishi Bommasani et al., *On the Opportunities and Risks of Foundation Models*, arXiv:2108.07258 (July 12, 2022).

[148] *See* Rishi Bommasani et al., *On the Opportunities and Risks of Foundation Models*, arXiv:2108.07258 (July 12, 2022); *See also* Krystyna Marcinek et al., *Risk-Based AI Regulation: A Primer on the Artificial Intelligence Act of the European Union*, RAND (Nov. 20, 2024), https://www.rand.org/pubs/research_reports/RRA3243-3.html#fnb7.

[149] *See* Rishi Bommasani et al., *On the Opportunities and Risks of Foundation Models*, arXiv:2108.07258 (July 12, 2022).

[150] *See* Krystyna Marcinek et al., *Risk-Based AI Regulation: A Primer on the Artificial Intelligence Act of the European Union*, RAND (Nov. 20, 2024), https://www.rand.org/pubs/research_reports/RRA3243-3.html#fnb7.

[151] *See* Krystyna Marcinek et al*., Risk-Based AI Regulation: A Primer on the Artificial Intelligence Act of the European Union*, RAND (Nov. 20, 2024), https://www.rand.org/pubs/research_reports/RRA3243-3.html#fnb7.

[152] *See* CONG. RSCH. SERV., INTRODUCTION TO TORT LAW, IF11291, at 1 (2023).

[153] *See* Rebecca Crootof, *The Internet of Torts*, 69 DUKE L.J. 583, 587–88 (2019).

[154] *See* Gregory Smith et al., *Liability for Harms from AI Systems*, RAND (2024), https://www.rand.org/pubs/research_reports/RRA3243-4.html.

Soft law—industry guidelines, voluntary codes, and technical best practices—has emerged as a pragmatic workaround. It offers speed, flexibility, and adaptability.[158] But it lacks teeth.[159] Without binding obligations or independent enforcement, soft law depends on the goodwill of the very entities it seeks to guide.[160] Worse, many soft law regimes are dominated by the largest AI developers, raising concerns about capture and self-serving standard-setting.[161] The result is a patchwork governance landscape with little external accountability and uneven adoption.[162]

---

[155] *See* Gregory Smith et al., *Liability for Harms from AI Systems*, RAND (2024), https://www.rand.org/pubs/research_reports/RRA3243-4.html.

[156] *See* Gregory Smith et al., *Liability for Harms from AI Systems*, RAND (2024), https://www.rand.org/pubs/research_reports/RRA3243-4.html.

[157] *See* Cruz v. Talmadge, 244 F. Supp. 3d 231 (D. Mass. 2017); Nilsson v. General Motors LLC, No. 3:18-cv-00471 (N.D. Cal. filed Jan. 22, 2018); Mark Geistfeld, *Strict Products Liability 2.0*, 14 J. TORT L. 403, 419–20 (2021).

[158] *See generally* Gary Marchant & Carlos Ignacio Gutierrez, *Soft Law 2.0: An Agile and Effective Governance Approach for Artificial Intelligence*, 24 MINN. J.L. SCI. & TECH. 375, 376–78, 390 (2023); Anna Jobin, Marcello Ienca & Effy Vayena, *The Global Landscape of AI Ethics Guidelines*, 1 NAT. MACH. INTELL. 389 (2019)
; *Ethical Guidelines for Trustworthy AI*, EUROPEAN COMM. (2019); Jessica Fjeld et al., *Principled Artificial Intelligence: Mapping Consensus in Ethical and Rights-Based Approaches to Principles for AI*, BERKMAN KLEIN CTR. RES. PUB. No. 2020-1 (2020).

[159] *See generally* Gary Marchant & Carlos Ignacio Gutierrez, *Soft Law 2.0: An Agile and Effective Governance Approach for Artificial Intelligence*, 24 MINN. J.L. SCI. & TECH. 375, 376–78, 390 (2023); Anna Jobin, Marcello Ienca & Effy Vayena, *The Global Landscape of AI Ethics Guidelines*, 1 NAT. MACH. INTELL. 389 (2019)
; *Ethical Guidelines for Trustworthy AI*, EUROPEAN COMM. (2019); Jessica Fjeld et al., *Principled Artificial Intelligence: Mapping Consensus in Ethical and Rights-Based Approaches to Principles for AI*, BERKMAN KLEIN CTR. RES. PUB. No. 2020-1 (2020).

[160] *See generally* Gary Marchant & Carlos Ignacio Gutierrez, *Soft Law 2.0: An Agile and Effective Governance Approach for Artificial Intelligence*, 24 MINN. J.L. SCI. & TECH. 375, 376–78, 390 (2023); Anna Jobin, Marcello Ienca & Effy Vayena, *The Global Landscape of AI Ethics Guidelines*, 1 NAT. MACH. INTELL. 389 (2019)
; *Ethical Guidelines for Trustworthy AI*, EUROPEAN COMM. (2019); Jessica Fjeld et al., *Principled Artificial Intelligence: Mapping Consensus in Ethical and Rights-Based Approaches to Principles for AI*, BERKMAN KLEIN CTR. RES. PUB. No. 2020-1 (2020).

[161] *See* Kevin Wei et al., *How Do AI Companies "Fine-Tune" Policy? Examining Regulatory Capture in AI Governance*, ARXIV:2410.13042 (Oct. 2024), https://arxiv.org/abs/2410.13042; Chris Meserole, *Soft Law as a Complement to AI Regulation*, BROOKINGS (Mar. 3, 2021), https://www.brookings.edu/articles/soft-law-as-a-complement-to-ai-regulation/.

[162] *See generally* Gary Marchant & Carlos Ignacio Gutierrez, *Soft Law 2.0: An Agile and Effective Governance Approach for Artificial Intelligence*, 24 MINN. J.L. SCI. & TECH. 375, 376–78, 390 (2023); Anna Jobin, Marcello Ienca & Effy Vayena, *The Global Landscape of AI Ethics Guidelines*, 1 NAT. MACH. INTELL. 389 (2019)

Each of these regimes—regulation, tort, and soft law—aims to manage AI's risks, but each falls short in a different dimension. Regulation lags behind innovation. Tort law struggles with fault attribution under complexity. Soft law lacks legitimacy and enforcement. What is needed is a new governance layer: one that combines the incentive alignment of liability, the adaptability of market mechanisms, and the institutional reliability of public law.[163] Artificial intelligence is reconfiguring the structure of human decision-making.[164] Governing it will require institutions capable not only of reacting to harm, but of absorbing, pricing, and shaping systemic risk under deep uncertainty.[165]

This Note proposes that reinsurance—long used to stabilize high-risk sectors such as nuclear energy, agriculture, and healthcare—can provide that layer for frontier AI.[166] A federal reinsurance program would catalyze the development of a private insurance market for high-risk AI systems, translating ambiguous hazards into priced liabilities and aligning developer incentives with public safety at scale.[167]

## IV. THE SOLUTION: FEDERALLY BACKED REINSURANCE
### A. The Logic of Insurance & Reinsurance

Artificial intelligence is a transformative leap with no perfect historical precedent, but legal and institutional history still offers guidance.[168] When

---

; *Ethical Guidelines for Trustworthy AI*, EUROPEAN COMM. (2019); Jessica Fjeld et al., *Principled Artificial Intelligence: Mapping Consensus in Ethical and Rights-Based Approaches to Principles for AI*, BERKMAN KLEIN CTR. RES. PUB. No. 2020-1 (2020).

[163] *See* Shauhin A. Talesh, *Insurance Law as Public Interest Law*, 4 UC IRVINE L. REV. 985, 993–998 (2014); *See also* Gary Marchant et al., *Governing Emerging Technologies Through Soft Law: Lessons for Artificial Intelligence*, 61 JURIMETRICS 1 (2021).

[164] *See* Mariano-Florentino Cuéllar, *A Common Law for the Age of Artificial Intelligence: Incremental Adjudication, Institutions, and Relational Non-Arbitrariness*, 119 COLUM. L. REV. 1773, 1781–82 (2019); Rishi Bommasani et al., *On the Opportunities and Risks of Foundation Models*, ARXIV:2108.07258 (July 12, 2022), https://arxiv.org/abs/2108.07258.

[165] Steven M. Shavell, *Liability for Harm versus Regulation of Safety*, 13 J. LEGAL STUD. 357, 358–60 (1984) (on institutional roles in pricing and deterring risk); *See generally* PHILIP E. TETLOCK, SUPERFORECASTING: THE ART AND SCIENCE OF PREDICTION (2015) (discussing the notion of "deep uncertainty" in literature about forecasting and epistemic limits).

[166] *See generally* John E. Gudgel, *Insurance as a Private Sector Regulator and Promoter of Security and Safety* (2022) (Ph.D. dissertation, George Mason Univ.) https://mars.gmu.edu/server/api/core/bitstreams/c89687f4-4a4f-42a6-a4f3-53dd498d5321/content.

[167] *See* Anat Lior, *Insuring AI: The Role of Insurance in Artificial Intelligence Regulation*, 35 HARV. J.L. & TECH. 468, 470–79 (2022); Kenneth S. Abraham & Daniel Schwarcz, *Courting Disaster*, 27 CONN. INS. L.J. 407 (2021); Tom Baker & Rick Swedloff, *Regulation by Liability Insurance: From Auto to Lawyers Professional Liability*, 60 UCLA L. REV. 1412 (2013); *But see* Kenneth S. Abraham & Daniel Schwarcz, *The Limits of Regulation by Insurance*, 98 IND. L.J. 739, 741–43 (2023) (exploring the potential and limits of insurance institutions in managing systemic technological risk).

[168] *See* Anat Lior, *Insuring AI: The Role of Insurance in Artificial Intelligence Regulation*, 35 HARV. J.L. & TECH. 468, 470–74 (2022).

electricity reshaped industry,[169] when nuclear power altered the strategic calculus of war and peace,[170] and when oil became a global economic cornerstone,[171] each carried profound risks that were not well understood at the outset.[172] AI may prove just as foundational—and just as dangerous.[173] It could remake markets, reorganize labor, and optimize global systems. But it could also trigger unanticipated harms, from opaque decision-making to catastrophic system failures.[174] In such a fast-moving context, the central question is: how does society regulate something that moves faster than our regulatory machinery?

One answer lies in insurance.[175] The insurance industry exists to do what regulation often struggles with: price risk under uncertainty.[176] Every policy represents an implicit judgment—what can go wrong, how often, and at what cost. This makes insurance more than just a financial hedge; it is a disciplinary mechanism.[177] If AI developers are required to carry liability insurance for their systems, they would become accountable not only to public regulators but to private underwriters. Insurers could shape behavior by denying coverage, adjusting premiums, or excluding risky practices—tools that are often more nimble than legal mandates.[178]

At present, however, the cyber and technology insurance markets are too thin to support this function.[179] Participation is limited, actuarial data are scarce,

---

[169] *See* CARL BENEDIKT FREY, THE TECHNOLOGY TRAP: CAPITAL, LABOR, AND POWER IN THE AGE OF AUTOMATION 189–222 (2019) (describing electricity's role in industrial transformation and labor realignment).

[170] *See* DANIEL YERGIN, THE QUEST: ENERGY, SECURITY, AND THE REMAKING OF THE MODERN WORLD 370–71 (2011) (discussing the tension between growing the nuclear power industry and preventing arms proliferation: nuclear energy's dual-use dilemma).

[171] *See generally* DANIEL YERGIN, THE PRIZE: THE EPIC QUEST FOR OIL, MONEY & POWER (1991) (narrating the geopolitics of oil and how it shaped the 20th century).

[172] *See generally* DAVID A. MOSS, WHEN ALL ELSE FAILS: GOVERNMENT AS THE ULTIMATE RISK MANAGER (Harvard Univ. Press 2004) (discussing historical government interventions to manage foundational risks such as nuclear power and financial collapse).

[173] *See generally* STUART RUSSELL, HUMAN COMPATIBLE: ARTIFICIAL INTELLIGENCE AND THE PROBLEM OF CONTROL (2019) (on existential and systemic risks posed by advanced AI); *See also* Executive Order No. 14,110, 88 Fed. Reg. 75191 (Oct. 30, 2023) (U.S. federal recognition of AI's transformative and risky nature).

[174] Rebecca Crootof, *The Internet of Torts*, 69 DUKE L. J. 583, 591–92 (2019).

[175] *See* Martin Eling, *How Insurance Can Mitigate AI Risks*, BROOKINGS (Nov. 7, 2019), https://www.brookings.edu/articles/how-insurance-can-mitigate-ai-risks/.

[176] *See* Shauhin A. Talesh, *Insurance Law as Public Interest Law*, 4 UC IRVINE L. REV. 985, 993–998 (2014).

[177] *See* Tom Baker & Rick Swedloff, *Regulation by Liability Insurance: From Auto to Lawyers Professional Liability*, 60 UCLA L. REV. 1412 (2013); *But see* Kenneth S. Abraham & Daniel Schwarcz, *The Limits of Regulation by Insurance*, 98 IND. L.J. 739, 741–43 (2023) (exploring the potential and limits of insurance institutions in managing systemic technological risk).

[178] *See* Cristian E. Trout, *Liability and Insurance for Catastrophic Losses: The Nuclear Power Precedent and Lessons for AI* (paper presented at the Generative AI and Law Workshop, Int'l Conf. on Mach. Learning, Vienna, Austria, July 2024), https://arxiv.org/pdf/2409.06672.

[179] *See* Kenneth S. Abraham & Daniel Schwarcz, *Courting Disaster*, 27 CONN. INS. L.J. 407 (2021) (warning about catastrophic cyber-risk and proposing new liability frameworks); But see *Governments Should Not Be the Cyber Insurers of Last Resort*, FIN. TIMES (Mar. 17, 2024),

and underwriting models remain immature.[180] The result is a self-reinforcing cycle: high premiums deter companies from buying coverage, which in turn limits data collection and prevents accurate risk modeling, keeping premiums high.[181] This feedback loop traps the market in a pre-institutional phase—underdeveloped, uncertain, and fragile.[182]

This is precisely where federal reinsurance can intervene. Reinsurance is a basic pillar of modern risk management, allowing primary insurers to offload part of their liability in exchange for a share of their premium income.[183] The logic is straightforward: when insurers know their exposure to catastrophic loss is capped, they are more willing to write policies in emerging or volatile markets. By assuming the tail-end of the risk curve, a federal reinsurance program for AI would enable insurers to price policies more competitively, thereby drawing more firms into the risk pool.[184]

Over time, deeper participation improves the quality of actuarial data, refines underwriting standards, and allows both public and private actors to map the risk landscape with greater fidelity.[185] This is not just market stabilization—it is governance through institutional learning.[186]

There are two principal forms of reinsurance: proportional and non-proportional. In proportional reinsurance, reinsurers share a fixed percentage of both premiums and losses, operating almost as co-underwriters.[187] In non-proportional (or excess-of-loss) reinsurance, the reinsurer pays only when claims exceed a certain threshold. This second model is especially well-suited for AI, where the goal is not

---

https://www.ft.com/content/b119fd0c-f0a4-4221-bb18-4dfc23a6d81c.https://www.ft.com/content/b119fd0c-f0a4-4221-bb18-4dfc23a6d81c

[180] *See* Kenneth S. Abraham & Daniel Schwarcz, *Courting Disaster*, 27 CONN. INS. L.J. 407 (2021); But see *Governments Should Not Be the Cyber Insurers of Last Resort*, FIN. TIMES (Mar. 17, 2024), https://www.ft.com/content/b119fd0c-f0a4-4221-bb18-4dfc23a6d81c.https://www.ft.com/content/b119fd0c-f0a4-4221-bb18-4dfc23a6d81c

[181] *See* Kenneth S. Abraham & Daniel Schwarcz, *Courting Disaster*, 27 CONN. INS. L.J. 407 (2021); But see *Governments Should Not Be the Cyber Insurers of Last Resort*, FIN. TIMES (Mar. 17, 2024), https://www.ft.com/content/b119fd0c-f0a4-4221-bb18-4dfc23a6d81c.https://www.ft.com/content/b119fd0c-f0a4-4221-bb18-4dfc23a6d81c

[182] *See* Kenneth S. Abraham & Daniel Schwarcz, *Courting Disaster*, 27 CONN. INS. L.J. 407 (2021); But see *Governments Should Not Be the Cyber Insurers of Last Resort*, FIN. TIMES (Mar. 17, 2024), https://www.ft.com/content/b119fd0c-f0a4-4221-bb18-4dfc23a6d81c.https://www.ft.com/content/b119fd0c-f0a4-4221-bb18-4dfc23a6d81c

[183] Niels Viggo Haueter, *Reinsurance Function and Marketlocked*, OXFORD RES. ENCYC. ECON. FIN. (Nov. 19, 2020), https://doi.org/10.1093/acrefore/9780190224851.013.268.

[184] Seth J. Carroll et al., *Assessing the Effectiveness of the Terrorism Risk Insurance Act*, RAND Corp. (2004), https://www.rand.org/pubs/research_briefs/RB9153.html.

[185] Cristian Trout, *Liability and Insurance for Catastrophic Losses: The Nuclear Power Precedent and Lessons for AI*, in GENERATIVE AI AND LAW WORKSHOP, INT'L CONF. ON MACHINE LEARNING (Vienna, Austria 2024).

[186] *See* Anat Lior, *Insuring AI: The Role of Insurance in Artificial Intelligence Regulation*, 35 HARV. J.L. & TECH. 468, 479 (2022); *Accord* Kenneth S. Abraham & Daniel Schwarcz, *Courting Disaster*, 27 CONN. INS. L.J. 1, 58 (2021).

[187] *See* Anat Lior, *Insuring AI: The Role of Insurance in Artificial Intelligence Regulation*, 35 HARV. J.L. & TECH. 468, 479 (2022).

to manage routine software bugs but to absorb the costs of low-probability, high-consequence failures—the "long-tail" events that define systemic technological risk.[188]

A federal reinsurance program for AI would do more than lower premiums. It would embed incentives for safety, tying insurability to risk management practices and transparency standards.[189] It would create a mechanism for managing catastrophic events without collapsing private markets. And perhaps most importantly, it would leverage the analytical capabilities of the insurance industry itself, offering a form of adaptive, data-driven oversight that can evolve alongside the technology it governs.[190]

## B. Four Institutional Precedents for Federal Reinsurance

*1. Nuclear Energy: The Price-Anderson Act*

The Price-Anderson Nuclear Industries Indemnity Act, originally passed in 1957 and subsequently amended, was a pragmatic solution to a problem that threatened to paralyze the U.S. nuclear power industry before it began.[191] Private insurers refused to underwrite nuclear plants—not because the technology lacked promise, but because the potential liabilities were vast, novel, and incalculable.[192] Unlike fires or automobile accidents, nuclear incidents lacked actuarial baselines; there was no reliable way to estimate either their frequency or the scale of damage they might produce.[193]

To resolve this impasse, Congress enacted Price-Anderson, establishing a three-tiered liability regime.[194] First, private insurers were required to provide a baseline amount of coverage for licensed reactors.[195] Second, the industry was compelled to contribute to a collective pool that would cover losses exceeding individual policy limits.[196] Finally, the federal government acted as a reinsurer of last resort, absorbing liabilities above the industry's aggregate cap.[197] The statute also provided exclusive federal jurisdiction for nuclear tort claims and established procedural standards to streamline litigation.[198]

---

[188] *See* Kenneth S. Abraham & Daniel Schwarcz, *Courting Disaster*, 27 CONN. INS. L.J. 1, 58 (2021).

[189] *See* Anat Lior, *Insuring AI: The Role of Insurance in Artificial Intelligence Regulation*, 35 HARV. J.L. & TECH. 468, 479 (2022).

[190] *See* Martin Eling, *How Insurance Can Mitigate AI Risks*, BROOKINGS (Nov. 7, 2019), https://www.brookings.edu/articles/how-insurance-can-mitigate-ai-risks/; See e.g. *Maritime Casualty Data Keeps World Trade Moving*, LLOYD'S LIST INTELLIGENCE (Feb. 23, 2024), https://www.lloydslistintelligence.com/thought-leadership/blogs/maritime-casualty-data-world-trade-moving.

[191] Price-Anderson Act, Pub. L. No. 85-256, 71 Stat. 576 (1957) (codified as amended at 42 U.S.C. § 2210 (2023)); *See generally* John E. Gudgel, *Insurance as a Private Sector Regulator and Promoter of Security and Safety* (2022) (Ph.D. dissertation, George Mason Univ.) https://mars.gmu.edu/server/api/core/bitstreams/c89687f4-4a4f-42a6-a4f3-53dd498d5321/content.

[192] U.S. Nuclear Regulatory Comm'n, *The Price-Anderson Act: 2021 Report to Congress* 2–5 (2021), https://www.nrc.gov/docs/ML2133/ML21335A064.pdf (describing the Act's three-tier liability system, federal jurisdiction, and procedural mechanisms for managing nuclear tort claims).

[193] U.S. Nuclear Regulatory Comm'n, *The Price-Anderson Act: 2021 Report to Congress* 2–5 (2021), https://www.nrc.gov/docs/ML2133/ML21335A064.pdf.

[194] *See* 42 U.S.C. § 2210(b)–(d) (detailing the three-layer liability regime)..

[195] 42 U.S.C. § 2210(b)–(d); U.S. Nuclear Regulatory Comm'n, *The Price-Anderson Act: 2021 Report to Congress* 2–5 (2021), https://www.nrc.gov/docs/ML2133/ML21335A064.pdf.

The Act served two primary functions: it ensured compensation for victims of nuclear incidents while also removing liability barriers to industry participation. In doing so, it created the conditions for commercial nuclear energy to develop under a regime of bounded, shared risk. It did not eliminate liability—it redistributed it, institutionalizing a legal infrastructure capable of managing tail events too extreme for private actors alone.[199]

Artificial intelligence now stands at a similar juncture. Like nuclear energy in the mid-20th century, AI is a general-purpose technology[200] with both transformative potential and catastrophic downside risk.[201] And like nuclear accidents, certain AI failures—particularly those involving large-scale misuse, misalignment, or autonomous systems—could produce consequences beyond the reach of conventional liability doctrines or private insurance capacity.[202]

A federal reinsurance program for AI, modeled on Price-Anderson's structure, would offer a layered solution: primary coverage from private insurers; a pooled industry fund for distributed, non-systemic claims; and a federal backstop for the rare but severe events that threaten broader societal harm.[203] While the substantive risks differ, the institutional challenge is the same: how to govern technological development under radical uncertainty. Price-Anderson succeeded not by perfecting predictive models, but by building a legal architecture capable of absorbing the worst-case scenario.[204] That remains the core design problem for AI governance today.

---

[196] 42 U.S.C. § 2210(b)–(d); U.S. Nuclear Regulatory Comm'n, *The Price-Anderson Act: 2021 Report to Congress* 2–5 (2021), https://www.nrc.gov/docs/ML2133/ML21335A064.pdf.
[197] 42 U.S.C. § 2210(b)–(d); U.S. Nuclear Regulatory Comm'n, *The Price-Anderson Act: 2021 Report to Congress* 2–5 (2021), https://www.nrc.gov/docs/ML2133/ML21335A064.pdf.
[198] 42 U.S.C. § 2210(b)–(d); U.S. Nuclear Regulatory Comm'n, *The Price-Anderson Act: 2021 Report to Congress* 2–5 (2021), https://www.nrc.gov/docs/ML2133/ML21335A064.pdf.
[199] *See* U.S. Nuclear Regulatory Comm'n, *The Price-Anderson Act: 2021 Report to Congress* 1–2 (2021), https://www.nrc.gov/docs/ML2133/ML21335A064.pdf (explaining that the Act aimed to compensate victims while facilitating industry participation through shared liability mechanisms).
[200] *See* Rishi Bommasani et al., *On the Opportunities and Risks of Foundation Models*, arXiv:2108.07258 (July 12, 2022), https://arxiv.org/abs/2108.07258 (describing AI as a general-purpose technology and identifying both its positive potential and systemic risks).
[201] *See* Gabriel Weil, *Tort Law as a Tool for Mitigating Catastrophic Risk from Artificial Intelligence* 5–10 (Jan. 2024), https://papers.ssrn.com/abstract=4694006 (arguing that traditional tort law may not suffice for catastrophic AI harms and exploring liability gaps); *Accord* Kenneth S. Abraham & Daniel Schwarcz, *Courting Disaster*, 27 CONN. INS. L.J. 407, 411–13 (2021) (discussing how catastrophic technological risk can exceed insurance market capacity, with implications for AI); *See generally* Chan et al., *Harms from Increasingly Agentic Algorithmic Systems* (2023).
[202] Gabriel Weil, *Tort Law as a Tool for Mitigating Catastrophic Risk from Artificial Intelligence* 5–10 (Jan. 2024), https://papers.ssrn.com/abstract=4694006. Kenneth S. Abraham & Daniel Schwarcz, *Courting Disaster*, 27 CONN. INS. L.J. 407, 411–13 (2021); Chan et al., *Harms from Increasingly Agentic Algorithmic Systems* (2023).
[203] Price-Anderson Act, Pub. L. No. 85-256, 71 Stat. 576 (1957) (codified as amended at 42 U.S.C. § 2210 (2023)); *See* U.S. Nuclear Regulatory Comm'n, *The Price-Anderson Act: 2021 Report to Congress* 2–5 (2021), https://www.nrc.gov/docs/ML2133/ML21335A064.pdf.

*2. Agriculture: Federal Crop Insurance*

Agriculture operates in fundamentally different risk environment. Farmers do not worry about sudden, civilization-scale catastrophes; they worry instead about persistent, cyclic threats: droughts, floods, pests and volatile commodity markets.[205] The losses are frequent and often predictable, but they remain financially destabilizing—especially when concentrated across regions or seasons.[206]

To manage this volatility, the federal government built the Federal Crop Insurance Program, overseen by the U.S. Department of Agriculture's Risk Management Agency.[207] Unlike reinsurance in nuclear energy, which functions primarily as a catastrophic backstop agricultural reinsurance is embedded in a hybrid public-private structure.[208] Farmers purchase policies from private insurers, but those insurers operate under a system of federal subsidies, underwriting standards, and reinsurance guarantees.[209] This partnership enables coverage in markets where repeated losses would otherwise drive insurers out entirely.[210]

The logic is simple but profound: by socializing some portion of risk, the government transforms an otherwise fragile insurance market into a planning infrastructure. Reinsurance allows insurers to remain solvent across bad years and good. In turn, it allows farmers to plant, borrow, and invest—despite the inevitability of loss.[211]

While the mechanics differ, the governing principle echoes that of the Price-Anderson Act: when private insurers face structural barriers to covering an essential but unstable sector, public intervention can stabilize the system without displacing market forces entirely.²¹² In this way, agricultural reinsurance offers a model not for catastrophic tail risk, but for routine, distributed uncertainty—a feature that may prove equally relevant in the broader landscape of artificial intelligence governance.²¹³

---

[204] Price-Anderson Act, Pub. L. No. 85-256, 71 Stat. 576 (1957) (codified as amended at 42 U.S.C. § 2210 (2023)); U.S. Nuclear Regulatory Comm'n, *The Price-Anderson Act: 2021 Report to Congress* 2–5 (2021), https://www.nrc.gov/docs/ML2133/ML21335A064.pdf..

[205] *See* U.S. Dep't of Agric., *Risk Management Agency: Crop Insurance Basics* 2–3 (2021), https://www.rma.usda.gov/Topics/Crop-Insurance-Basics (describing the predictable but financially destabilizing risks faced by farmers, including weather variability and price volatility).

[206] *See* U.S. Dep't of Agric., *Risk Management Agency: Crop Insurance Basics* 2–3 (2021), https://www.rma.usda.gov/Topics/Crop-Insurance-Basics.

[207] 7 U.S.C. §§ 1501–1524 (2023); *See generally* U.S. Dep't of Agric., Risk Mgmt. Agency, https://www.rma.usda.gov/. (last visited Mar. 25, 2025).

[208] 7 U.S.C. §§ 1501–1524 (2023); *See* U.S. Dep't of Agric., *Risk Management Agency: Crop Insurance Basics* 2–3 (2021), https://www.rma.usda.gov/Topics/Crop-Insurance-Basics (describing the predictable but financially destabilizing risks faced by farmers, including weather variability and price volatility).

[209] 7 U.S.C. §§ 1501–1524 (2023); U.S. Dep't of Agric., *Risk Management Agency: Crop Insurance Basics* 2–3 (2021), https://www.rma.usda.gov/Topics/Crop-Insurance-Basics.

[210] 7 U.S.C. §§ 1501–1524 (2023); U.S. Dep't of Agric., *Risk Management Agency: Crop Insurance Basics* 2–3 (2021), https://www.rma.usda.gov/Topics/Crop-Insurance-Basics.

[211] U.S. Gov't Accountability Off., *Crop Insurance: Opportunities Exist to Improve Program Delivery and Reduce Costs* (GAO-17-501, 2017), https://www.gao.gov/products/gao-17-501. (noting that federal support enables insurers to remain solvent during high-loss years, sustaining market participation and producer confidence).

*3. Medicine: Malpractice and the Affordable Care Act*

Healthcare occupies a middle ground between farming and nuclear energy. Most medical procedures are routine and predictable, but some cases—like malpractice suits or catastrophic diagnoses—create financial volatility that private insurers struggle to absorb.²¹⁴ Reinsurance helps insurers manage this volatility by spreading high-dollar losses across larger risk pools.²¹⁵

The Affordable Care Act (ACA) created a federal reinsurance program under 42 U.S.C. § 18061 to stabilize the individual insurance market by reimbursing insurers for high-cost enrollees.²¹⁶ The logic was straightforward: if insurers were shielded from the full cost of the sickest patients, they would be less likely to avoid them. But the policy's implementation raised critical legal and administrative questions.²¹⁷ In *Health Republic Insurance Co. v. United States* and *Blue Cross & Blue Shield of North Carolina v. United States*, the Court of Federal Claims held that the federal government had to honor unpaid risk corridor reimbursements promised under the ACA.²¹⁸ These rulings underscore a key principle: federal reinsurance mechanisms are only credible if they provide predictable, legally enforceable backstops.

---

²¹² See Price-Anderson Nuclear Industries Indemnity Act, 42 U.S.C. § 2210 (2023).
²¹³ See U.S. Nuclear Regulatory Comm'n, *The Price-Anderson Act: 2021 Report to Congress* 2–5 (2021), https://www.nrc.gov/docs/ML2133/ML21335A064.pdf (explaining how the Act enabled nuclear development by addressing insurance market failure through layered public-private risk sharing); *See also* U.S. Gov't Accountability Off., *Crop Insurance: Opportunities Exist to Improve Program Delivery and Reduce Costs* (GAO-17-501, 2017), https://www.gao.gov/products/gao-17-501.
; Rishi Bommasani et al., *On the Opportunities and Risks of Foundation Models*, arXiv:2108.07258 (July 12, 2022), https://arxiv.org/abs/2108.07258 (noting that the risks posed by AI range from localized failures to systemic harms across domains).
²¹⁴ U.S. Gen. Acct. Off., *Medical Malpractice Insurance: Multiple Factors Have Contributed to Increased Premium Rates*, GAO-03-702, at 15 (2003), https://www.gao.gov/assets/gao-03-702.pdf.
²¹⁵ U.S. Gen. Acct. Off., *Medical Malpractice Insurance: Multiple Factors Have Contributed to Increased Premium Rates*, GAO-03-702, at 15–27 (2003), https://www.gao.gov/assets/gao-03-702.pdf.
²¹⁶ 42 U.S.C. § 18061(b)(4); *Transitional Reinsurance Program*, CTR. FOR MEDICARE & MEDICAID SERVS., https://www.cms.gov/marketplace/health-plans-issuers/premium-stabilization-programs/transitional-reinsurance-program (last visited Mar. 26, 2025).
²¹⁷ 42 U.S.C. § 18061(b)(4); *Transitional Reinsurance Program*, CTR. FOR MEDICARE & MEDICAID SERVS., https://www.cms.gov/marketplace/health-plans-issuers/premium-stabilization-programs/transitional-reinsurance-program (last visited Mar. 26, 2025).
²¹⁸ Health Republic Ins. Co. v. United States, 129 Fed. Cl. 757, 772–73 (2017); Blue Cross & Blue Shield of N.C. v. United States, 131 Fed. Cl. 125, 130 (2017).

Other litigation raised questions about the structure and scope of federal reinsurance. In *New Mexico Health Connections v. United States HHS*, insurers challenged methodologies used in the ACA's risk adjustment program.[219] In *Ohio v. United States*, states objected to federal mandates regulating their insurance markets.[220] Together, these cases reveal a fundamental tension: federal reinsurance can stabilize private markets, but only if designed with clear statutory authority, sustainable funding, and procedural transparency.

Reinsurance also plays a vital role in the context of medical malpractice. Catastrophic claims—birth injuries, surgical errors, wrongful death—can bankrupt smaller insurers. Reinsurance absorbs the tail-end risk, but debates continue over the role of public authority.[221] In *Gerhart v. United States HHS*, the court considered how much federal oversight is appropriate for state-run reinsurance pools.[222] Meanwhile, regulations such as 45 C.F.R. § 800.204[223] establish solvency standards for multi-state plans, while cases like *Conway v. United States*[224] and *Richardson v. United States*[225] explore the boundaries of public liability under the Federal Tort Claims Act (FTCA).[226] These precedents collectively demonstrate that medical reinsurance is not just an economic device—it is a governance institution, one that requires careful calibration between federal oversight and private-sector innovation.

What does this mean for artificial intelligence? A well-designed AI reinsurance program should combine elements from all three sectors. Like agriculture, AI failures may be frequent but non-catastrophic, demanding risk pooling and routine coverage.[227] Like medicine, AI liability will likely require complex legal frameworks, blending federal standards with state-level discretion.[228] And like nuclear energy, the most extreme AI scenarios—model misalignment, emergent capabilities, or catastrophic misuse—will require a federal backstop to absorb losses too large for the private sector to bear.[229]

---

[219] New Mexico Health Connections v. U.S. Dep't of Health & Hum. Servs., 946 F.3d 1138, 1142–45 (10th Cir. 2019).
[220] Ohio v. United States, 154 F. Supp. 3d 621, 627–30 (S.D. Ohio 2016).
[221] U.S. Gen. Acct. Off., *Medical Malpractice Insurance: Multiple Factors Have Contributed to Increased Premium Rates*, GAO-03-702, at 15 (2003), https://www.gao.gov/assets/gao-03-702.pdf.
[222] Gerhart v. U.S. Dep't of Health & Hum. Servs., 242 F. Supp. 3d 827, 834–36 (N.D. Iowa 2017).
[223] 45 C.F.R. § 800.204.
[224] Conway v. United States, 647 F.3d 228, 232–34 (D.C. Cir. 2011).
[225] Richardson v. United States, 841 F.2d 993, 996–98 (9th Cir. 1988).
[226] Federal Tort Claims Act, ch. 753, 60 Stat. 842 (1946) (codified as amended at 28 U.S.C. §§ 1346(b), 2671–2680).
[227] *See* U.S. Gov't Accountability Off., *Crop Insurance: Opportunities Exist to Improve Program Delivery and Reduce Costs* (GAO-17-501, 2017), https://www.gao.gov/products/gao-17-501.
[228] *See* Gerhart v. U.S. Dep't of Health & Hum. Servs., 242 F. Supp. 3d 827, 834–36 (N.D. Iowa 2017); 45 C.F.R. § 800.204 (2023); Conway v. United States, 647 F.3d 228, 232–34 (D.C. Cir. 2011).
[229] *See* Price-Anderson Act, Pub. L. No. 85-256, 71 Stat. 576 (1957) (codified as amended at 42 U.S.C. § 2210 (2023)); U.S. NUCLEAR REG. COMM'N, THE PRICE-ANDERSON ACT: 2021 REPORT

The challenge is to design a reinsurance architecture that is financially viable, legally robust, and institutionally flexible—one that encourages innovation while preparing for worst-case scenarios.[230] The ACA and medical malpractice models show that this is possible, but they also warn us: without reliable funding mechanisms, enforceable guarantees, and clarity around risk attribution, even well-intentioned reinsurance schemes can collapse under the weight of litigation and political pressure.[231] The AI context raises these stakes exponentially.[232]

*4. Finance: Building an architecture of confidence*

The financial sector offers a final instructive model. Like artificial intelligence, modern finance is an extremely complex,[233] opaque,[234] and inter-reliant system.[235] So, failures in one corner can cascade through the entire network.[236] The 2008 financial crisis revealed how risk pooling, opacity in modeling, and underpriced tail events could produce catastrophic outcomes.[237] In response, the federal government reaffirmed its role as *insurer of last resort*—not only for depositors through the FDIC,[238] but for broader financial institutions through mechanisms like

---

TO CONGRESS 1–5 (2021), https://www.nrc.gov/docs/ML2133/ML21335A064.pdf (describing the Act's federal indemnity for nuclear incidents); *See generally* Rishi Bommasani et al., *On the Opportunities and Risks of Foundation Models*, arXiv:2108.07258 (July 12, 2022), https://arxiv.org/abs/2108.07258 (identifying catastrophic AI risks including misalignment and emergent behavior).

[230] *See* Health Republic Ins. Co. v. United States, 129 Fed. Cl. 757, 772–73 (2017); Blue Cross & Blue Shield of N.C. v. United States, 131 Fed. Cl. 125, 130 (2017); Gerhart v. U.S. Dep't of Health & Hum. Servs., 242 F. Supp. 3d 827, 834–36 (N.D. Iowa 2017); U.S. Gov't Accountability Off., *Crop Insurance: Opportunities Exist to Improve Program Delivery and Reduce Costs* (GAO-17-501, 2017), https://www.gao.gov/products/gao-17-501.

[231] *See* Health Republic Ins. Co. v. United States, 129 Fed. Cl. 757, 772–73 (2017); Blue Cross & Blue Shield of N.C. v. United States, 131 Fed. Cl. 125, 130 (2017); Gerhart v. U.S. Dep't of Health & Hum. Servs., 242 F. Supp. 3d 827, 834–36 (N.D. Iowa 2017); U.S. Gov't Accountability Off., U.S. Gov't Accountability Off., *Crop Insurance: Opportunities Exist to Improve Program Delivery and Reduce Costs* (GAO-17-501, 2017), https://www.gao.gov/products/gao-17-501.

[232] See Rishi Bommasani et al., *On the Opportunities and Risks of Foundation Models*, arXiv:2108.07258 (July 12, 2022), https://arxiv.org/abs/2108.07258; Gabriel Weil, *Tort Law as a Tool for Mitigating Catastrophic Risk from Artificial Intelligence* 5–10 (Jan. 2024), https://papers.ssrn.com/abstract=4694006.

[233] *See* FIN. CRISIS INQUIRY COMM'N, *The Financial Crisis Inquiry Report,* 1-3 (2011), https://www.congress.gov/112/chrg/CHRG-112hhrg64556/CHRG-112hhrg64556.pdf

[234] *See* FIN. CRISIS INQUIRY COMM'N, *The Financial Crisis Inquiry Report,* 32-34 (2011), https://www.congress.gov/112/chrg/CHRG-112hhrg64556/CHRG-112hhrg64556.pdf

[235] *See* FIN. CRISIS INQUIRY COMM'N, *The Financial Crisis Inquiry Report,* 36-38 (2011), https://www.congress.gov/112/chrg/CHRG-112hhrg64556/CHRG-112hhrg64556.pdf

[236] *See generally* FIN. CRISIS INQUIRY COMM'N, *The Financial Crisis Inquiry Report* (2011), https://www.congress.gov/112/chrg/CHRG-112hhrg64556/CHRG-112hhrg64556.pdf

[237] *See generally* FIN. CRISIS INQUIRY COMM'N, *The Financial Crisis Inquiry Report* (2011), https://www.congress.gov/112/chrg/CHRG-112hhrg64556/CHRG-112hhrg64556.pdf

[238] Federal Deposit Insurance Act, 12 U.S.C. § 1821 (2023).

the Troubled Asset Relief Program (TARP)[239] and the Federal Reserve's emergency lending facilities.[240]

Of particular relevance here, is the FDIC's structure. Created by the Banking Act of 1933,[241] the FDIC insures deposits up to a statutory limit,[242] funded by risk-adjusted premiums paid by participating banks.[243] It is not merely a backstop—it disciplines behavior by pricing risk, evaluating bank health,[244] and exercising oversight through examinations and resolution planning.[245] This model—government-backed, industry-funded, and actuarially managed—offers a robust example of *confidence infrastructure*: a system designed not just to prevent failure, but to preserve trust and continuity under stress.[246]

Artificial intelligence, especially frontier model deployment, raises similar concerns.[247] Where finance concentrates economic risk, AI may concentrate informational and decision-making risk.[248] Just as financial regulators struggled to map systemic exposure to risk across counterparties and synthetic instruments, AI regulators may struggle to track how foundation models propagate risk across sectors and jurisdictions.[249]

---

[239] Emergency Economic Stabilization Act of 2008, Pub. L. No. 110-343, 122 Stat. 3765 (establishing TARP); *See also* U.S. Dept. Treas., *Troubled Asset Relief Program (TARP)*, https://home.treasury.gov/policy-issues/financial-markets-financial-institutions-and-fiscal-service/tarp.

[240] *See* U.S. Gov't Accountability Off., *Federal Reserve Lending Programs: Status of Monitoring and Main Street Lending Program*, GAO-24-106482 (2023), https://www.gao.gov/assets/d24106482.pdf; FIN. STAB. OVERSIGHT COUNCIL, 2023 ANNUAL REPORT (2023), https://home.treasury.gov/system/files/261/FSOC2023AnnualReport.pdf; BOARD OF GOVS. FED. RES. SYS., 110TH ANNUAL REPORT (2023), https://www.federalreserve.gov/publications/files/2023-annual-report.pdf.

[241] *See* Banking Act of 1933, Pub. L. No. 73-66, § 8, 48 Stat. 162, 168 (codified as amended at 12 U.S.C. §§ 1811–1835a (2023)).

[242] 12 U.S.C. § 1821(a)(1)(A).

[243] 12 U.S.C. § 1817(b) (2023); FDIC, *A Brief History of Deposit Insurance* (1998), https://www.fdic.gov/publications/brief-history-deposit-insurance; *See* Edward Garnett et al., *A History of Risk-Based Premiums at the FDIC,* FDIC Staff Study No. 2020-01 (2020), https://www.fdic.gov/analysis/cfr/staff-studies/2020-01.pdf; *See also* FDIC, *Resolution Plans Required for Insured Depository Institutions with $100 Billion or More in Total Assets; Informational Filings Required for Insured Depository Institutions with at Least $50 Billion but Less Than $100 Billion in Total Assets*, 89 Fed. Reg. 56620 (July 9, 2024), https://www.fdic.gov/board/final-rule-12-cfr-36010-federal-register-notice.

[244] *See* Adam J. Levitin, *Safe Banking: Finance and Democracy*, 83 U. CHI. L. REV. 357, 378–92 (2016); *See also* Edward Garnett et al.**,** *A History of Risk-Based Premiums at the FDIC*, FDIC Staff Study No. 2020-01, at 10–20 (2020), https://www.fdic.gov/analysis/cfr/staff-studies/2020-01.pdf.

[245] Resolution Plans Required for Insured Depository Institutions With $100 Billion or More in Total Assets, 89 Fed. Reg. 56620 (July 9, 2024), https://www.fdic.gov/board/final-rule-12-cfr-36010-federal-register-notice.

[246] Jonathan R. Macey & Geoffrey P. Miller, *Deposit Insurance, the Implicit Regulatory Contract, and the Mismatch in the Term Structure of Banks' Assets and Liabilities*, 12 YALE J. REG. 1, 17–22 (1995); GAO, *Deposit Insurance: Assessment of Regulators' Use of Prompt Corrective Action Provisions* (GAO-07-242, Feb. 2007), https://www.gao.gov/assets/gao-07-242.pdf.

[247] Risto Uuk et al., *A Taxonomy of Systemic Risks from General-Purpose AI*, arXiv 2–4 (2024), https://arxiv.org/abs/2412.07780.

[248] Markus Anderljung et al., *Frontier AI Regulation: Managing Emerging Risks to Public Safety*, arXiv 3–4 (2023), https://arxiv.org/abs/2307.03718.

A federal AI reinsurance program could borrow from the FDIC in three ways: (1) it could require participation for high-risk developers,[250] (2) it could calibrate premiums to safety and transparency practices,[251] and (3) it could accumulate institutional expertise for managing model failure and risk spillover.[252] The goal, as in finance, is not merely to respond to crises after they happen—but to create an ecosystem where trust, accountability, and solvency are maintained in advance.

**C. Applying the Reinsurance Model to Artificial Intelligence**

Again, reinsurance is not a new idea. It is the reason farmers can survive bad harvests, nuclear reactors operate with congressional blessing, how hospitals stay solvent despite million-dollar mistakes, and how banks can extend enough credit for a complex economy.

But every industry demands a different model. Agriculture deals with frequent but constant losses—droughts, floods, pests. Government-backed crop insurance helps farmers stay afloat. In nuclear energy, the stakes are higher. Accidents are rare but catastrophic. Liability caps and multi-layered insurance pools keep the industry from collapsing under the weight of a worst-case scenario. Medicine sits in between, balancing routine liability with the risk of catastrophe, so insurers rely on a mix of private coverage, government programs, and legal protections. Finance offers another useful example: deposit insurance and capital rules are designed to stop small problems from turning into full-blown crises.

Artificial intelligence may need a model that borrows from all of the above. Routine failures, such as bias in training data or misclassifications, will be frequent.[253] Liability will be diffuse and complex, often implicating both developers and end-users.[254] And frontier models, the most powerful, unpredictable, and general-purpose systems, may someday trigger tail events so extreme that they exceed the capacity of both courts and insurers to manage on their own.[255]

---

[249] Rishi Bommasani et al., *On the Opportunities and Risks of Foundation Models*, arXiv 19 (2021), https://arxiv.org/abs/2108.07258.
[250] 12 U.S.C. § 1815(a) (2023); *See* Adam J. Levitin, *Safe Banking: Finance and Democracy*, 83 U. Chi. L. Rev. 357, 384–85 (2016).
[251] 12 U.S.C. § 1817(b)(1)(C) (2023).
[252] Resolution Plans Required for Insured Depository Institutions with $100 Billion or More in Total Assets, 89 Fed. Reg. 56620 (July 9, 2024), https://www.fdic.gov/board/final-rule-12-cfr-36010-federal-register-notice.
[253] *See generally* U.S. Nat'l Inst. of Standards & Tech., *AI Risk Management Framework* 13–16 (2023) https://nvlpubs.nist.gov/nistpubs/ai/NIST.AI.600-1.pdf.
[254] *See* Gabriel Weil, *Tort Law as a Tool for Mitigating Catastrophic Risk from Artificial Intelligence* 5–10 (Jan. 2024).
[255] *See* Rishi Bommasani et al., *On the Opportunities and Risks of Foundation Models*, arXiv:2108.07258 (July 12, 2022), https://arxiv.org/abs/2108.07258 (describing foundation models' frequent failure modes and potential for systemic harm); *See* Gabriel Weil, *Tort Law as a Tool for Mitigating Catastrophic Risk from Artificial Intelligence* 5–10 (Jan. 2024).

The insurance industry is already grappling with these issues.[256] Lloyd's of London has warned that AI blurs traditional liability categories.[257] Who pays when a self-driving car crashes? Not the driver, who was not driving.[258] Not the AI itself, which is not a legal entity—for now. That leaves software suppliers and manufacturers.[259] As AI expands into fields like diagnostics, investment advising, and legal decision-making, similar gaps emerge.[260] Professional liability doctrines stretch. Product liability doctrines creak.[261] The result is a system with mounting uncertainty and no consistent framework for allocating AI-driven risk.[262]

In the near term, insurers will adapt through contractual instruments: indemnification clauses, performance warranties, and bespoke policies.[263] But these tools only work in known contexts.[264] For frontier AI systems—models that exhibit emergent capabilities or potentially unaligned goals[265]—insurers lack both historical data and pricing mechanisms.[266] They face not actuarial risk, but epistemic uncertainty.[267] What they need is a structure that can price the unpriceable, at least in the aggregate.

---

[256] *See generally* LLOYD'S, GENERATIVE AI: TRANSFORMING THE CYBER LANDSCAPE (2023), https://www.lloyds.com/news-and-insights/futureset/futureset-insights/generative-ai-transforming-the-cyber-landscape.

[257] *See generally* LLOYD'S, TAKING CONTROL: ARTIFICIAL INTELLIGENCE AND INSURANCE, EMERGING RISK REPORT 2019-TECHNOLOGY, https://assets.lloyds.com/assets/pdf-taking-control-aireport-2019-final/1/pdf-taking-control-aireport-2019-final.PDF.

[258] *See* Erin Mindoro Ezra, Rebecca L. Rogers & Robert D. Chan, *Life on Autopilot: Self-Driving Cars Raise Liability and Insurance Questions and Uncertainties*, REUTERS (Aug. 30, 2024), https://www.reuters.com/legal/legalindustry/life-autopilot-self-driving-cars-raise-liability-insurance-questions-2024-08-30/.

[259] *See* TAKING CONTROL: ARTIFICIAL INTELLIGENCE AND INSURANCE, EMERGING RISK REPORT 2019-TECHNOLOGY, LLOYD'S 36

[260] *See* TAKING CONTROL: ARTIFICIAL INTELLIGENCE AND INSURANCE, EMERGING RISK REPORT 2019-TECHNOLOGY, LLOYD'S 39; See also *Artificial Intelligence*, LLOYD'S MARKET ASS. (2023), https://www.lmalloyds.com/LMA/Hot_topics/LMA/Hot_Topics/Artificial_Intelligence.aspx

[261] *See* Morgan Sansone *Motor Vehicle Accidents Caused by Autonomous Vehicles: Exploring AI Liability in the Tort System* ARIZ. STATE L. J. (Mar. 23, 2023), https://arizonastatelawjournal.org/2023/03/23/motor-vehicle-accidents-caused-by-autonomous-vehicles-exploring-ai-liability-in-the-tort-system/ (discussing the need for adaptable liability frameworks as AI technologies evolve).

[262] See *Insurers Will Not Find It Easy to Share the Road with Self-Driving Cars*, FIN. TIMES, (Jan. 15, 2025), https://www.ft.com/content/53aa25aa-ca33-4a67-b315-8802376639ef.

[263] *See* Anat Lior, *Insuring AI: The Role of Insurance in Artificial Intelligence Regulation*, 35 HARVARD J.L. & TECH. 389, 395–97 (2022); *Liability for Harms from AI Systems: The Application of U.S. Tort Law to Artificial Intelligence* RAND 30–32 (2024), https://www.rand.org/pubs/research_reports/RRA3243-4.html.

[264] *See* Anat Lior, *Insuring AI: The Role of Insurance in Artificial Intelligence Regulation*, 35 HARVARD J.L. & TECH. 389, 395–97 (2022).

[265] *See* Rishi Bommasani et al., *On the Opportunities and Risks of Foundation Models*, arXiv:2108.07258 (July 12, 2022), https://arxiv.org/abs/2108.07258

[266] *See* Anat Lior, *Insuring AI: The Role of Insurance in Artificial Intelligence Regulation*, 35 HARVARD J.L. & TECH. 389, 395–97 (2022), https://ssrn.com/abstract=4266259.

[267] *See* AMLIN & OXFORD MARTIN SCH., SYSTEMIC RISK OF MODELLING IN INSURANCE: DID YOUR MODEL TELL YOU ALL MODELS ARE WRONG? 11–14 (2015); Anat Lior, *Innovating Liability: The*

# V. THE PROPOSAL: A TRIPARTITE FRAMEWORK FOR AI RISK TRANSFER

This Note proposes a three-tiered liability and insurance framework to govern the deployment of frontier artificial intelligence (AI) systems. Drawing on historical precedents in nuclear energy, agriculture, medicine, and finance, the framework adapts longstanding risk governance tools to the unique epistemic and systemic risks posed by powerful, general-purpose models. Each tier is designed to match a specific layer of the AI risk landscape—from frequent, distributed failures to catastrophic, existential-scale events.

## A. Part One: Mandatory Private Insurance for Frontier AI Developers

The first tier requires AI developers working with frontier models—defined by capability, scale, or compute thresholds, such as those outlined in California's SB 1047[268]—to carry private liability insurance as a condition of deployment. This mirrors the compulsory coverage required for activities that pose high but uncertain risks, such as medical malpractice or hazardous industrial operations. Private insurers would evaluate developers' safety practices, model transparency, and security controls as underwriting criteria. This would embed a market-based accountability mechanism into the AI development pipeline, encouraging best practices before systems are released.

This design aligns with recent work in AI governance that emphasizes the need for *verifiable claims* about the properties and development conditions of AI systems. As Brundage et al. argue, high-level ethical commitments are insufficient to guarantee safety, security, or fairness. What is needed are concrete institutional mechanisms—such as third-party auditing, red teaming, and structured transparency obligations—that enable both developers and external stakeholders to substantiate system-level claims. This proposal incorporates those same mechanisms, not as stand-alone governance tools, but as underwriting criteria within a liability insurance ecosystem. In doing so, it transforms verification from a normative aspiration into a financial necessity. Insurers, acting as market-aligned evaluators, assume the functional role of third-party auditors, embedding accountability into the deployment pipeline without requiring ex ante regulatory mandates or universal compliance regimes.[269]

## B. Part Two: An Industry-Funded Risk Pool for Non-Catastrophic Losses

The second tier introduces an industry-funded pool to stabilize the market for routine but unpredictable losses. Like Pool Re (the United Kingdom's terrorism

---

*Virtuous Cycle of Torts, Technology and Liability Insurance*, 25 YALE J.L. & TECH. 448, 460–62 (2023).
[268] S.B. 1047, 2023–2024 Leg., Reg. Sess. (Cal. 2024) (enrolled Sept. 3, 2024), https://leginfo.legislature.ca.gov/faces/billNavClient.xhtml?bill_id=202320240SB1047.
[269] Miles Brundage et al., *Toward Trustworthy AI Development: Mechanisms for Supporting Verifiable Claims* 8–15 (Apr. 2020), https://arxiv.org/abs/2004.07213.

reinsurance facility)[270] or the U.S. Federal Crop Insurance Program,[271] this structure would absorb correlated claims that exceed individual insurers' expectations but fall short of systemic collapse. AI-specific examples might include widespread business interruption from model hallucination, reputational harms from biased outputs, or supply chain losses from embedded AI decision failures.[272] The pooled layer would promote actuarial learning, reduce premium volatility, and enable insurers to remain solvent even during concentrated risk events.

Realizing this tier, however, depends on shared epistemic infrastructure—a common baseline for identifying, recording, and evaluating model-related harms—provided by the first tier. As Brundage et al. emphasize, effective AI governance demands not just ethical aspiration but mechanisms that make claims about safety, fairness, and reliability verifiable.[273] Their proposed tools—such as the structured sharing of AI incidents, the use of red teaming and bias bounties, and the standardization of audit documentation—correspond to the first tier and serve as the reporting and verification backbone for the pooled risk layer. Reinsurers and industry pools alike would benefit from these mechanisms as inputs to collective underwriting, enabling a dynamic yet accountable learning system across firms.[274]

## C. Part Three: A Federal Reinsurance Backstop for Catastrophic Loss

Finally, the third tier provides a federal reinsurance backstop to cover the tail-end of AI risk—rare but devastating events where private capacity vanishes. Analogous to the *Price-Anderson Act* in nuclear energy[275] and the *Terrorism Risk Insurance Act (TRIA)*,[276] this federal layer would activate in the event of system-wide alignment failures, adversarial deployment of foundation models, or self-propagating harm from autonomous agents.[277] By absorbing these extreme losses, the federal government would stabilize the private insurance market, maintain developer accountability, and ensure compensability even in black-swan scenarios.[278]

---

[270] *Government-Guaranteed Insurance Against Systemic Risk: Pool Re,* OFFICE FOR BUDGET RESPONSIBILITY, (Nov. 2022), https://obr.uk/box/government-guaranteed-insurance-against-systemic-risk-pool-re/.

[271] S*ee generally* U.S. GOV'T ACCOUNTABILITY OFF., GAO-17-501, CROP INSURANCE: OPPORTUNITIES EXIST TO IMPROVE PROGRAM DELIVERY AND REDUCE COSTS (2017).

[272] *See* Rishi Bommasani et al., *On the Opportunities and Risks of Foundation Models* 43–45 (July 12, 2022), https://arxiv.org/abs/2108.07258.

[273] Miles Brundage et al., *Toward Trustworthy AI Development: Mechanisms for Supporting Verifiable Claims* 8–15 (Apr. 2020), https://arxiv.org/abs/2004.07213.

[274] See Miles Brundage et al., *Toward Trustworthy AI Development: Mechanisms for Supporting Verifiable Claims* 19–21 (Apr. 2020), https://arxiv.org/abs/2004.07213.

[275] S*ee* Price-Anderson Act, Pub. L. No. 85-256, 71 Stat. 576 (1957) (codified as amended at 42 U.S.C. § 2210 (2023)).

[276] *See* Terrorism Risk Insurance Act of 2002 (TRIA), Pub. L. No. 107-297, 116 Stat. 2322 (codified at 15 U.S.C. §§ 6701 note, 6721–6728 (2023)).

[277] *See* Rishi Bommasani et al., *On the Opportunities and Risks of Foundation Models* 114–16 (July 12, 2022), https://arxiv.org/abs/2108.07258.

[278] *See generally* Kenneth S. Abraham & Daniel Schwarcz, *Courting Disaster*, 27 CONN. INS. L.J. 407 (2021).

Precedents underscore the feasibility of a federally backed reinsurance framework with minimal public expenditure. Under the *Price-Anderson Nuclear Industries Indemnity Act*, established in 1957 to manage liability for nuclear incidents, the nuclear insurance pools have paid approximately $151 million in claims over several decades, averaging about $2.5 million per year. Notably, these costs have been borne by the private sector, requiring *$0 in federal payouts*. Similarly, the Terrorism Risk Insurance Act (TRIA), enacted in 2002 to stabilize the insurance market post-9/11, has facilitated a public-private partnership where the federal government shares losses from certified acts of terrorism exceeding certain thresholds. As of now, the federal government has not made significant payouts under TRIA, as no terrorist events triggering the coverage thresholds have occurred since its enactment. These models demonstrate that well-structured reinsurance programs can provide substantial coverage for catastrophic risks while imposing negligible recurring costs on the federal government. [279]

### D. Legal Structure and Administrative Feasibility

The reinsurance program could be housed within an existing federal risk authority—such as the Federal Insurance Office[280] or a new AI Risk Management Agency—with oversight mechanisms tied to the National Institute of Standards and Technology (NIST).[281] Actuarial models, underwriting standards, and eligibility criteria would evolve over time, using data collected from the first and second tiers to refine pricing and exclusions.[282] The program would function not as a regulatory substitute but as an institutional complement—a flexible layer of governance that prices the unpriceable, disciplines risky behavior, and maps the emerging landscape of AI hazards.[283]

## CONCLUSION

This structure is not just a financial patch. It is a governance tool. Insurance governs by exclusion: what cannot be underwritten often cannot be deployed. It governs by pricing: riskier systems carry higher premiums or require mitigation plans. And it governs by information: the underwriting process demands

---

[279] *See* Energy Contractors Price-Anderson Grp., Response to U.S. Dep't of Energy Notice of Inquiry on Preparation of Report to Congress on The Price-Anderson Act, (Oct. 25, 2021); U.S. DEP'T OF THE TREAS., REPORT ON THE EFFECTIVENESS OF THE TERRORISM RISK INSURANCE PROGRAM, 89 FR 19639 (2024).

[280] S*ee* Dodd–Frank Wall Street Reform and Consumer Protection Act, Pub. L. No. 111-203, § 502, 124 Stat. 1376, 1580 (2010) (codified at 31 U.S.C. § 313 (2023)); U.S. DEP'T OF THE TREASURY, FEDERAL INSURANCE OFFICE, ANNUAL REPORT 1–2 (2023).

[281] S*ee* Exec. Order No. 14,110, 88 Fed. Reg. 75,193, 75,197 (Oct. 30, 2023); Cal. S.B. 1047, 2023–2024 Leg., Reg. Sess. § 22605(a)(2) (Cal. 2024).

[282] *See* Anat Lior, *Insuring AI: The Role of Insurance in Artificial Intelligence Regulation*, 35 HARV. J.L. & TECH. 389, 396–97 (2022); Kenneth S. Abraham & Daniel Schwarcz, *Courting Disaster*, 27 CONN. INS. L.J. 407 (2021).

[283] *See* Anat Lior, *Insuring AI: The Role of Insurance in Artificial Intelligence Regulation*, 35 HARV. J.L. & TECH. 389, 396–97 (2022); *See generally* Kenneth S. Abraham & Daniel Schwarcz, *Courting Disaster*, 27 CONN. INS. L.J. 407 (2021)

documentation, scenario analysis, and disclosure. In a field as opaque as AI, these pressures are not secondary—they are foundational.

Moreover, even a narrow federal reinsurance program—one that applies only to the most powerful and dangerous systems—can have systemic effects. Insurers gain experience modeling high-risk AI. Developers seeking insurability adopt standardized practices. Information generated for frontier underwriting bleeds into mid-tier applications. What begins as a subsidy becomes an epistemic infrastructure: a map of the risk landscape that other actors—regulators, researchers, litigants—can use.

Reinsurance will not solve AI governance. But it may be the only way to start managing catastrophic risk at scale—through incentives, not dictates; through pricing, not prohibition; through institutional design, not moral panic.

This Note has argued that a federally backed reinsurance program for high-risk AI systems offers a scalable and institutionally coherent solution to the alignment problem's policy dimension. It would not only stabilize insurance markets but also generate risk intelligence, align incentives, and support an ecosystem of oversight. If designed carefully—modeled on historical precedents and tailored to the architecture of modern AI development—it could become a foundational element of AI governance in the twenty-first century.

History does not reward passivity. It rewards preparation. As our society approaches this new technological ocean, we would do well to remember that the compass and the sextant did not guide ships across dangerous waters alone. It was the contract, the risk pool, and the promise of shared responsibility that made the voyage possible. Reinsurance may now do for AI what it once did for empire and industry: transform uncertainty into direction, and risk into strategy.